\documentclass[12pt]{article}

\usepackage[utf8]{inputenc}
\usepackage{roboto}

\usepackage{setspace}
\usepackage{lscape}
\usepackage[bottom=1in,top=1in,left=1in,right=1in]{geometry}
\usepackage{indentfirst}
\usepackage{titlesec}
\titlespacing*{\chapter}{0pt}{-15pt}{40pt}
\titlespacing{\section}{0pt}{0cm}{0cm}
\titlespacing{\subsection}{0pt}{0cm}{0cm}
\usepackage{appendix}

\usepackage{multirow}

\usepackage{enumerate}
\usepackage[shortlabels]{enumitem}
\setlist[itemize]{topsep=0pt}
\setlist[enumerate]{topsep=0pt}

\usepackage{colortbl}
\usepackage[dvipsnames]{xcolor}
\definecolor{blue}{RGB}{0,0,153}
\definecolor{red}{RGB}{204,0,0}
\definecolor{yellow}{RGB}{255,255,153}
\definecolor{green}{RGB}{0,102,0}
\definecolor{orange}{RGB}{255,128,0}
\definecolor{purple}{RGB}{102,0,204}
\definecolor{orange_graph}{RGB}{245, 121, 58}
\definecolor{blue_graph}{RGB}{15, 32, 128}

\newcolumntype{a}{>{\columncolor{blue!15}}c}

\usepackage{amssymb,amsfonts,amsthm}
\usepackage[fleqn]{amsmath}
\usepackage{mathtools}
\usepackage{bbm}

\usepackage{graphicx,graphics,url,caption}
\usepackage{subcaption} 
\usepackage{float}
	
\usepackage{tikz}
\usetikzlibrary{decorations.pathreplacing,angles,quotes}
\usetikzlibrary{patterns}
\usepackage{array,booktabs,arydshln,xcolor}

\usepackage{adjustbox}
\usepackage[flushleft]{threeparttable}

\usepackage{natbib}
\setcitestyle{citesep={,}}

\usepackage{siunitx}
\newcolumntype{d}{S[input-symbols = ()]}

\usepackage{listings}
\usepackage{color} 
\definecolor{mygreen}{RGB}{28,172,0} 
\definecolor{mylilas}{RGB}{170,55,241}

\newcommand\fnsep{\textsuperscript{,}}

\usepackage{clipboard}
\newclipboard{EER_RR}
\usepackage{xr-hyper}
\usepackage{hyperref}


\begin{document}

\title{\vspace{-1cm}\textbf{Labor market conditions and college graduation: evidence from Brazil}\footnote{I would like to thank Joseph Altonji, Costas Meghir, Cormac O'Dea, Jaime Arellano-Bover, Seth Zimmerman, Eduardo Ferraz, and participants of the Labor/Public Finance seminars at Yale University for their comments. I also owe thanks to Andre Portela, Rodrigo Oliveira, and Alei dos Santos for their help with the data. This project received IRB approval from the Yale Human Research Protection Program (2000030142) and the Research Ethics Committee, led by Fundação Getulio Vargas (P.050.2021).}}
\author{Lucas Finamor\footnote{Yale University --- Department of Economics. Contact: lucas.finamor@yale.edu.}}

\date{\small First Draft: January 2022\\[.2cm]This Draft: March 2023}

\maketitle\thispagestyle{empty}

\pagenumbering{arabic}  

\begin{abstract}
\noindent College students graduating in a recession have been shown to face large and persistent negative effects on their earnings, health, and other outcomes. This paper investigates whether students delay graduation to avoid these effects. Using data on the universe of students in higher education in Brazil and leveraging variation in labor market conditions across time, space, and chosen majors, the paper finds that students in public institutions delay graduation to avoid entering depressed labor markets. {A typical recession causes the on-time graduation rate to fall by 6.5\% in public universities and there is no effect on private institutions. The induced delaying increases average graduation by 0.11 semesters, consistent with 1 out of 18 students delaying graduation by one year in public universities.} The delaying effect is larger for students with higher scores, in higher-earnings majors, and from more advantaged backgrounds. This has important implications for the distributional impact of recessions. 
\end{abstract}

\textbf{JEL Codes:} I23,I24,J24,J21

\doublespacing

\clearpage
\pagenumbering{arabic} 
\section{Introduction}

College students face strong and persistent adverse effects when graduating in a recession. A recent survey by \cite{vonwachter2020} finds that, on average, college students graduating in a recession earn 10\% less, an effect that persists for ten years following graduation.\footnote{\cite{arellano2020} reviews papers exploring the graduating-in-a-recession effect, showing documented evidence from Austria, Belgium, Britain, Canada, Finland, Japan, Korea, Netherlands, Norway, Spain, and the United States of America.} \cite*{altonji2016} and \cite{arellano2020b} attribute part of these effects to students graduating in a recession finding their first jobs in lower-paying occupations or with smaller firms. \cite{forsythe2020} documents that hiring rates fall faster for young workers during recessions. The adverse effects of graduating during a recession are not limited to labor market outcomes: cohorts graduating in recessions experience worse outcomes on health, family formation, and crime \citep{vonwachter2020}. 

This raises the question of whether college students postpone graduation to avoid entering a depressed labor market. While there are direct and opportunity costs of delaying graduation, these costs could be outweighed, for some students, by avoiding the scarring effect of unemployment and finding a better match in the labor market later on. Universities facilitate networking opportunities and provide infrastructure to help job-seeking students, which can be more valuable with higher labor demand. Additionally, maintaining status as a student can be beneficial since they can enjoy subsidies and internship opportunities and increase human capital by attending more courses. 

Whether (and for whom) the benefits outweigh the costs is, ultimately, an empirical question. In this paper, I investigate if the local labor market conditions affect college students' graduation decisions in Brazil. By investigating which students avail of opportunities to delay graduation, I highlight a new dimension of heterogeneity in the costs of recessions.

To answer these questions, I bring together a rich collection of data. My primary data set consists of longitudinal data for the universe of students enrolled in any higher education program in Brazil since 2009, the Higher Education Census. For every student, I have demographic information, characteristics of the chosen major and institution, and, critically, the expected and actual graduation dates. Using national matched employer-employee data, I construct a measure for the labor market conditions specific for each major and state, which I term the major-weighted hires (MWH). To construct it, I retrieve the number of hires for each occupation and state from the matched employer-employee data. The final measure is the average of hires, weighted by the importance of each occupation for each major. Like common measures of the labor market conditions, such as \emph{local} unemployment rates, MWH explores labor market variations across time (students expected to graduate at different years) and geography (students graduating in different states).\footnote{Since majors have different expected durations (from 3 to 6 years), even controlling for cohort, I can still explore residual variation in expected graduation time.} However, MWH provides unique detail by major; students from the same state with the same expected graduation date may have different labor market opportunities, depending on how the typical occupations for their majors are trending. 

Exploring variation in employment conditions at the expected graduation date, I show that the overall effect of labor market conditions on graduating decisions is indistinguishable from zero. However, this aggregate effect masks substantial heterogeneity, particularly by the type of university. Students from public universities are less likely to graduate on time when facing a weaker labor market. Reducing the \emph{weighted} hiring by 1\% implies that these students are 0.07pp less likely to graduate on time. In the 2014--2016 recession in Brazil, the weighted hiring fell by 30\% on average, implying that the on-time graduation rate for public students was 2.1pp lower.\footnote{The recession in 2014--2016 raised the unemployment rate by 5.5pp in Brazil. A recession that increases the unemployment rate by 4-6pp is the typical recession analyzed in the literature evaluating the effects of graduating in a recession.} This effect represents a 6.5\% reduction in graduation at the expected time, translating to an increase in the average graduation time by around 0.11 semesters. That is equivalent to 1 out of 18 students (5.5\%) delaying graduation by one year.\footnote{In terms of the unemployment rate, I estimate the on-time graduation falling 0.2--0.4pp for an increase of 1pp in the unemployment rate.}

In contrast, I do not find any response on the timing of graduation for students in private institutions. These contrasting results are not surprising given the institutional differences between public and private institutions. Differences in course quality and students' characteristics can only account for part of the gap in the public-private estimates. Therefore, it is likely that public institutions being tuition-free plays a role in explaining these results. Postponing graduation for students in private institutes is significantly more costly. This makes the results from this paper even more relevant for several countries where the majority of the tertiary system is free or at a relatively low cost. 

In terms of heterogeneity, the effects of a recession on postponing graduation are more pronounced for students in higher-earnings majors and better socioeconomic status. Students with more advantaged backgrounds have more family resources to rely on while postponing labor market entry. In particular, this suggests a channel through which the educational system might foster inequalities -- more privileged students are better able to shield themselves from the adverse effects of labor market fluctuations by more freely choosing when to graduate. 

I complement this analysis by gathering data from one large public university in the state of Bahia --- The Federal University of Bahia (UFBA). UFBA is the state's flagship university, admitting 4,200 students annually. This data has information on credit accumulation and entry scores and can be linked with the matched employer-employee data. The credit accumulation is only responsive to the labor market conditions in the semesters near expected graduation, indicating that students may reduce their course load to postpone graduation. The delaying effect comes entirely from students without jobs. Students that had a formal job the year before graduation do not change their graduation decisions. The delaying effect is higher for students with higher entry scores in the admission exams and better socioeconomic status. 

Several papers document the negative effects on college students graduating in a recession \citep{genda2010,kahn2010,oreopoulos2012,altonji2016,schwandt2019,arellano2020b,rothstein2021}. This paper contributes to this literature by showing that some students react to labor market conditions by delaying graduation. There is abundant evidence that the choice to enter college is responsive to labor market conditions.\footnote{For example, \cite{betts1995,card2001,petrongolo2002,raaum2006,clark2011,hershbein2012,barr2013,sievertsen2016,stuart2020}.} Here, I explore a different margin, showing how students in the final years of college still respond to the labor market conditions by adjusting the time of graduation and, therefore, when to fully join the labor market. So, even if some individuals do not change their educational attainment, they still respond to the environment by adjusting graduation timing.\footnote{{A smaller literature also documents how graduate school enrollment responds to business cycle fluctuations \citep{bedard2008,johnson2013}. Due to data limitations, I cannot investigate this in this paper. However, I do not believe this is a relevant margin of adjustment for students graduating from college in Brazil. The proportion of college graduates with post-graduate degrees is very small, around 6--7\%, according to the Demographic Census in 2010. The Census data also indicate that a significant fraction of graduate school enrollment is not immediately after graduation, which eases the concerns for this analysis.}}\label{pag:grad_school_margin}

The heterogeneous effects complement the findings and supplement our understanding of existing literature. \cite{genda2010,oreopoulos2012,altonji2016,schwandt2019} and \cite{arellano2020b} find stronger negative scarring effects for students in lower-paying majors and with lower socioeconomic status. This is consistent with my findings that there is a smaller delaying effect for students with less advantaged backgrounds. My results reinforce the inequality concerns about who bears the costs of recessions raised by these papers. 

My paper relates to other papers that evaluate the relationship between labor market characteristics and late graduation. \cite{chen2016} shows that students more pessimistic about the labor market are more likely to plan for late graduation. \cite{bozick2009} uses a 2003--2004 survey in the US to show how some colleges accommodate students in times of a depressed labor market. \cite{messer2010} uses a sample of Swiss graduates from 1981--2001 to show how higher unemployment leads to a lower time-to-degree. {\cite{pechacek2013} explores state-level variation in unemployment to assess whether a sample of US college students from 1995 and 2003 are more likely to delay graduation.}\label{pag:juliepechacek} Lastly, \cite{aina2020} explores a sample of students graduating in 2002--2003 from 24 universities in Italy, showing that unemployment can be associated with late graduation and that this delay is costly for students. Relative to this work, my paper offers several contributions. First, my data covers the universe of college students in Brazil from more than 2,300 higher education institutes and 40,800 programs expected to graduate in 22 semesters. This universal coverage is beneficial to estimating the effect for a more representative sample and is essential to gauge the heterogeneity analysis. The different results depending on the type of university, majors, and individual characteristics are paramount to understanding the college-market transition and the effects of recessions. Second, my sample does not rely on students graduating in a given year, making it possible to control for cohort and still explore the effects of different labor market conditions.\footnote{Which, in addition, it allows the inference procedure to not impose strong assumptions, such as the non-correlation within geography or major over time.} Lastly, this paper also highly benefits from the matched employer-employee data, which allows me to compute labor market measures that vary not only across time and space but also across majors. 

Moreover, my results provide an explanation for \cite{kahn2010} and \cite{arellano2020}'s findings of larger effects when instrumenting the labor market conditions by the unemployment rate of expected graduation than in the OLS specifications. The compliers in their instrumental variables approach are precisely those not delaying graduation. I show that these students who do not delay graduation are more likely to be in lower-earnings majors and less advantaged backgrounds.

The large difference in the effect of recession on delaying graduation between public and private institutions demonstrates the key role of the institutional setting. These results align with \cite{garibaldi2012} and \cite{brunello2003}. {This points to one caveat: these results draw from the Brazilian institutional setting, which differs in important ways from the setting of several countries where the literature has documented the scarring effects.}\label{pag:brazilian_setting}

This paper proceeds as follows. I present the institutional setting of higher education in Brazil (Section 2); discuss the benefits and costs of delaying graduation (Section 3); present the data sets (Section 4); discuss the empirical strategy and the construction of the labor market measure (Section 5); present and discuss the empirical findings (Section 6); and offer concluding remarks (Section 7). 

\section{Institutional Setting}\label{institutional_setting}

As in many countries, admission to higher education in Brazil is institute and major-specific, commonly based solely on scores from admission exams. When students are admitted to a higher education institute, they are associated with a given major and a schedule depending on the period of classes: morning, afternoon, morning and afternoon, or evening. In this article, I will always refer to the major-institution-schedule as a \emph{program}. 

Higher education institutes can be universities, colleges, or college-centers (\emph{Centros Universitários}), depending on the range of degrees and majors offered. I do not distinguish between them and use the terms institutes or universities interchangeably. However, I make a key distinction between public and private institutes. Public institutions can be administered by federal, state, or municipal governments and are typically large research institutes. They do not charge any tuition and tend to have high-quality programs. Private institutes can be for-profit or nonprofit organizations, charge tuition, and are, on average, of lower quality than their public counterparts. Tuition in private institutes varies by major and institution. In 2017, the average monthly tuition for a business major was 262 USD (ranging from \$57 to \$1,502), while the average monthly tuition for a medicine major was 2,010 USD (\$1,095 to \$3,879).\footnote{To the best of my knowledge, there is no systematic collection of tuition data in the country. These numbers are from a survey conducted by a student guide publication (\emph{Guia do Estudante}). The numbers from 2017 are available at this link: \url{https://web.archive.org/web/20211118174614/https://guiadoestudante.abril.com.br/universidades/quanto-custa-fazer-uma-faculdade/}}\fnsep\footnote{Using a conversion of 1USD to 3.30 BRL in June of 2017.} In 2017, the average monthly per capita household income in the country was 389 USD.\footnote{1,285 BRL, using the household survey PNADC (\emph{Pesquisa Nacional por Amostra de Domíclios Contínua}).} Since public institutions are tuition-free and of superior quality, it is not surprising that they are highly selective. During my sample period, admission at public institutes had, on average, 12.1 candidates per seat, compared to only 1.6 in private institutions.\footnote{These numbers were drawn from the summary statistics for higher education produced by the INEP agency, linked to the Ministry of Education. For each year, I compute the number of applicants divided by the number of seats. The 12.1 and 1.6 are the average for this ratio for public and private institutions from 2009 to 2019.} Public universities account for roughly one-quarter of all college students. 

Programs' duration is given by the time students would take to graduate following the recommended course schedule. I will always refer to this time as the expected duration. Each program has a different duration, typically between 3 to 6 years. Variations in program duration can be primarily attributed to majors. For instance, Economics is typically a 4-year program, Law 5-years, and Medicine 6-years. Nevertheless, there is still some variation within majors. For example, 70\% of students in Economics are in 4-year programs, with the 30\% remaining in 4.5, 5, and 5.5-year programs.\footnote{Using the Higher Education Census. Please check Figure~\ref{fig:major_length} for some examples.} There can even be variation in the length of a program within a major at the same institution. For instance, the major of Economics at the University of Sao Paulo has an expected duration of 4 years if the student is enrolled in the morning schedule but 5 years if the student is enrolled in the evening schedule. The number of credits and the courses are exactly the same for the two schedules.\footnote{Source for the morning schedule: \url{https://web.archive.org/web/20211207164935/https://www.fea.usp.br/economia/graduacao/estrutura-curricular/diurno}. Source for the evening schedule: \url{https://web.archive.org/web/20211207165432/http://www.fea.usp.br/economia/graduacao/estrutura-curricular/noturno}}

\begin{figure}[!ht]
 \centering
 \caption{Distribution of time of graduation relative to expected graduation}\label{fig:distribution_grad}
 \includegraphics[width=\textwidth]{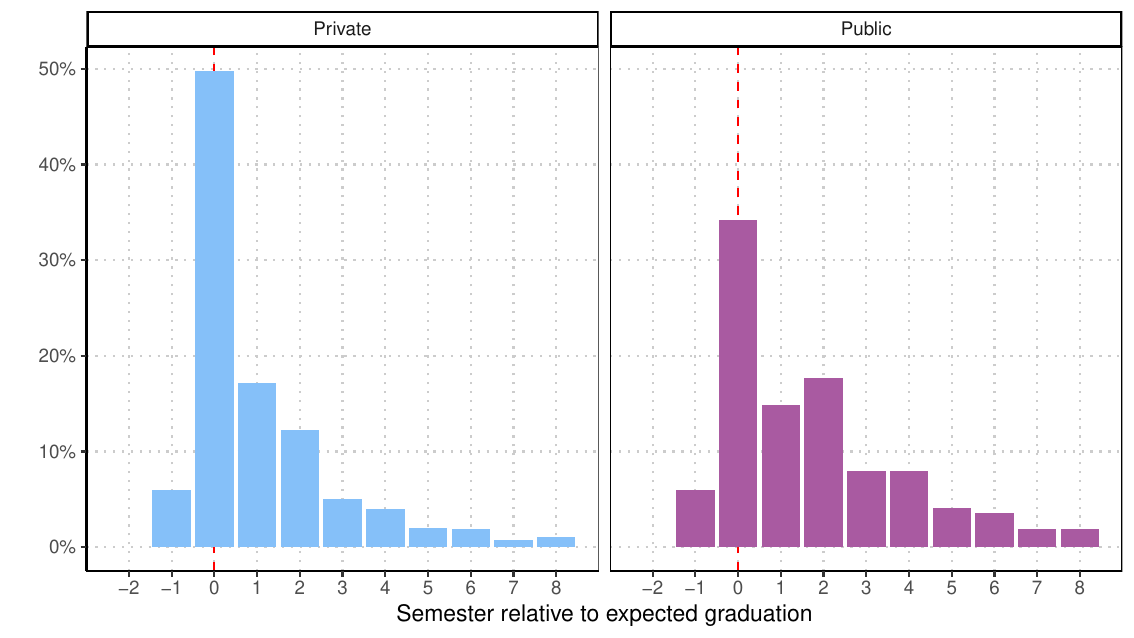}
 \caption*{\footnotesize{Notes: The figure presents the proportion of students that graduate each semester relative to the semester of expected graduation. The panel on the left shows the distribution for students in private institutions, while the panel on the right is for students in public institutions. I restrict the figure to those students graduating between -1 and 8 semesters relative to the correct time and enrolled in the first semester of their expected graduation year.}}
\end{figure}

While the numbers above refer to the expected duration of each program, students can take much longer to graduate. Figure \ref{fig:distribution_grad} shows the distribution of students graduating each semester relative to expected graduation.\footnote{I consider all students that graduate between -1 and 8 semesters off from their expected graduation, which accounts for almost the totality of students graduating.} We can see that 50\% of students in private institutions graduate at the expected time, while only 34\% do the same in public institutions. While on-time graduation is the most common time to graduate for all institutions, more than half of students do not graduate in the expected semester. These numbers are similar to those found in several European countries as reported by \cite{brunello2003,garibaldi2012,aina2018,aina2020}.

\begin{figure}[!ht]
 \centering
 \caption{Time trends of college enrollment in Brazil}\label{fig:time_trends}
 \begin{subfigure}[t]{0.49\linewidth}
 \caption{First Year Enrolled Students}\label{fig:ts_enrolle}
 \includegraphics[width=\textwidth]{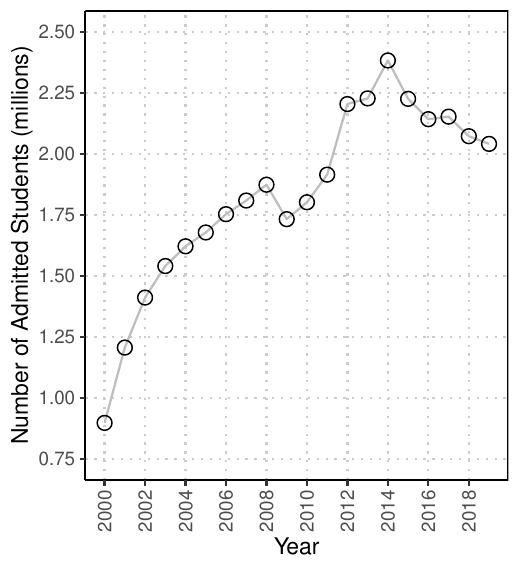}
 \end{subfigure}
 \begin{subfigure}[t]{0.49\linewidth}
 \caption{Proportion of 18yo in College}\label{fig:ts_prop18}
 \includegraphics[width=\textwidth]{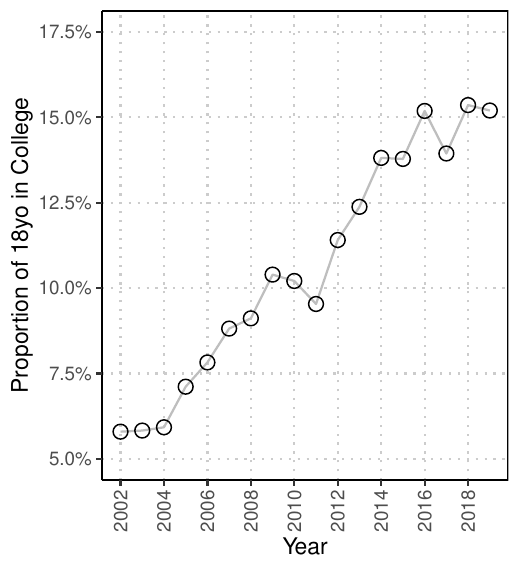}
 \end{subfigure}
 \caption*{\footnotesize{Notes: The panel on the left shows the number of admitted students in higher education across time. The panel on the right shows the proportion of 18 years-old individuals enrolled in any higher education program. The number of first-year enrolled students comes from the summary statistics produced by the INEP agency linked to the Ministry of Education. The proportion of 18 years old enrolled in higher education was computed using the household surveys (PNAD for 2002-2009 and 2011, Demographic Census for 2010, and PNADC for 2012-2019).}}
\end{figure}

Figure \ref{fig:time_trends} shows time trends on college enrollment in Brazil in the $21^{st}$ century. In panel (a), we can see that the number of students enrolled in the first year rose from 0.9 million to 2.4 million in 2014 and decreased to 2.0 million in 2019. The growth is due mainly to public policy at the federal level that created several new public higher education institutions and expanded programs providing scholarships and student loans for private institutions. These programs were severely affected by the 2014--2016 crisis and thus drastically reduced. In the second panel, we can see that this expansion was not merely due to population growth. The proportion of 18 years old students enrolled in higher education jumped from 5.8\% to more than 15\%. 

It is worth highlighting two additional aspects of the education system in Brazil. First, graduate programs are not a popular choice after obtaining a bachelor's degree. Only around 6.5\% and 2.0\% of those with bachelor's degrees have Masters and Doctoral degrees.\footnote{Using data from the 2020 Population Census.} Second, while public universities are of better quality and highly selective in higher education, the opposite occurs in primary and secondary education. Public primary and high schools are of inferior quality compared to their private counterparts, serving students from less advantaged backgrounds.\footnote{In the Appendix Figure \ref{fig:education_setting} I show the distribution of standardized scores for students in public and private schools for 5th, 9th, and 12th-graders, as well as for college students. Approximately 85\% of primary and secondary students are in public schools, with worse average scores than students in private schools. In Higher Education, we see the opposite. Less than 25\% of students are in public universities, and they exhibit higher average scores.} In the heterogeneity analysis, I will use the type of high school (public or private) as one of the proxies for socioeconomic status. 

\section{Graduation and the Labor Market Conditions}\label{sec:theoretical_framework}

Existing literature documents the negative and persistent effects of graduating in a recession. {While, to the best of my knowledge, there are no estimates for Brazil, the literature has shown robust effects in different settings. \cite{arellano2020} shows documented evidence for 11 countries with very different institutional settings.}\label{pag:evidence_brazil} More recent work has shown that initial placement is a critical driver of these negative results \citep{altonji2016,arellano2020}. Therefore, students near graduation could consider extending their college experience by delaying graduation and avoiding entering a depressed labor market. There are both costs and benefits associated with this decision.

Postponing graduation can be costly, as students forgo the labor earnings they would receive upon finding a job after graduating. There are additional direct monetary costs when staying enrolled, namely the tuition and other costs when in college (housing, materials, and others). It is also possible that postponing graduation is a bad signal for firms.

In terms of benefits, students could avoid the scarring effect of unemployment since it is difficult to find jobs when graduating in a recession.\footnote{The scarring effect can also include the psychological effects or the social pressure of being out of the university and unemployed.} An extra semester of college studies may look better to future employees, especially when unemployment is the counterfactual. Additionally, the university's infrastructure and network may be a more valuable resource as more active is the labor market. Students can also take the extra time to complete more credits and increase their human capital. 
Moreover, students are only eligible for internships when they are in school. Lastly, students have subsidies for transportation fares, food (in some public institutions), and cultural activities.

\begin{table}[h!]
    \centering
    \caption{Summary of costs and benefits associated with delaying graduation}
    \label{tab:costs_benefits}
    \begin{tabular}{ll}
        \hline \hline \\[-.3cm]
        \multicolumn{2}{l}{\textbf{Costs}} \\ \hline \\[-.3cm]
         \textbf{(C1)} & Forgone earnings while in college \\
         \textbf{(C2)} & Direct costs of attending college (tuition and others) \\
         \textbf{(C3)} & Worse signal for firms in the job-searching \\
         & \\
        \multicolumn{2}{l}{\textbf{Benefits}} \\ \hline \\[-.3cm]
         \textbf{(B1)} & Avoid the scarring effects of unemployment \\ 
         \textbf{(B2)} & Universities job-finding resources more valuable with higher demand \\
         \textbf{(B3)} & Complete additional coursework \\
         \textbf{(B4)} & Remaining eligibility for internships \\
         \textbf{(B5)} & Access to students' subsidies (e.g. transportation, cultural activities) \\
         \hline \hline 
    \end{tabular}
\end{table}

Table \ref{tab:costs_benefits} summarizes the discussed costs and benefits. My empirical strategy does not isolate the effect of each cost and benefit. However, the existing literature shows how several of these components can be important. \cite{garibaldi2012} shows how direct costs (C2) are highly relevant: lower tuition increases students' time to complete the degree. \cite{malacrino2017} show that the market values the number of credits taken and not only whether the degree was obtained, implying that the opportunity of completing additional courses (B3) is not negligible. \cite{nunley2016} also shows that internships are valuable in the job search, attributing option value to staying in college (B4).

We can undoubtedly expect highly heterogeneous effects since it is likely that those benefits and costs vary a lot from student to student and with their family's resources. The institutional setting presented in the last section hints that we should expect different results for students in public and private institutes. First, public schools do not charge tuition, making their students' direct costs (C2) orders of magnitude lower. On average, public universities are larger, of higher quality, and more prestigious, suggesting that the value of the university resources (B2), additional courses (B3), and the opportunity for additional internships (B4) are larger for them. Given that students in public intuitions have better scores and higher earnings potential, the tradeoff between avoiding the scarring (B1) and forgone earnings (C1) may be pronounced for them. Lastly, students in public universities have access to many more subsidies than their private counterparts, making (B5) arguably larger for them. Therefore, I hypothesize that the effects of delaying graduation are larger for students in public universities. 

\section{Data}

The primary data set is the Higher Education Census, containing annual microdata for every student enrolled in a higher education institute in Brazil from 2009 to 2019.\footnote{The Higher Education Census (\emph{Censo da Educação Superior}) is collected by the INEP agency (\emph{Instituto Nacional de Estudos e Pesquisas Educacionais Anísio Teixeira}), linked to the Ministry of Education.} For each student, the set includes information on the admission year, major, institute, and status (enrolled, on leave, or graduated). There is also basic demographic information, including date and place of birth, gender, and race. The level of observation is at the enrollment level, that is, a student-institute-major combination. 
From 2009 to 2017, it is possible to track students over time; however, starting in 2018, the individual identifiers are year-specific. Using a matching algorithm combining demographics and major-institution information, I link students over time, generating a panel from 2009 to 2019.\footnote{The matching procedure matches students by gender, date of birth, place of birth, major, university, and admission date, which is sufficient to uniquely identify individuals 95\% of the time. This procedure is detailed in Appendix \ref{data_details}.}\fnsep\footnote{In the Appendix Table~\ref{res:rob_2017} I show that the results are robust to only considering the data until 2017, where the unique identifiers are presented.} I then restrict the data set, keeping students from in-person bachelor-equivalent programs who are between 17 and 22 years of age when admitted and who have expected graduation dates before 2019. The resulting sample has 7.8 million unique individuals enrolled in 40,849 different programs, which are a combination of 74 majors, 2,342 higher education institutes, and 4 different schedules (morning, afternoon, morning-afternoon, and evening). Appendix \ref{data_details} contains more details about the data manipulation and sample selection.

Most of the analysis is based on a subsample of students enrolled in their expected graduation year. This is a sample of interest because I am testing whether students that could graduate decide to postpone graduation when facing worse labor market conditions. My analysis does not speak to the dynamics of the first years in college. This sampling definition is similar to the one by \cite{oreopoulos2012}. This restriction also maximizes the number of cohorts in the analysis. As the data only starts in 2009, the inclusion of all students enrolled in their first year would only allow me to track students entering in 2009 (with expected graduation dates between 2011 and 2014); whereas by restricting to the students enrolled in their final year, I can use all the cohorts with expected graduation dates between 2009 and 2019.\footnote{{Appendix~\ref{sample_restriction} shows how I would obtain qualitatively similar results without imposing this restriction and considering all students enrolled in the first semester of the program. Figure \ref{fig:robustness} also shows how the main results are robust to controlling for employment conditions for the four years before college graduation.}}\label{pag:sample_restriction} 

The Higher Education Census is valuable because of its breadth, covering the entire population of students in Brazil. However, it lacks more detailed information, such as credit accumulation, admission scores, and socioeconomic information at the individual level. To complement the analysis, I obtain from the \emph{Universidade Federal da Bahia} (UFBA), a large public university in the state of Bahia in Brazil, transcripts data from all students admitted between 2003 and 2017. The data contains the admission year, major, and courses enrolled in each semester for every student, including their course outcome (pass, fail, or dropped out) and their scores. I can also access the questionnaire that some students submit during the admissions process, which asks about students' demographics and socioeconomic standing. UFBA is the largest university in the state of Bahia, admitting about 4,200 students per year. With these data, I construct a panel of all students admitted between 2003 and 2015 and expected graduation in 2005-2017. I also apply the same restrictions I applied to the Higher Education Census sample, looking only at students in their last year, as detailed in Appendix \ref{data_details}.

I complement the information on student majors and institutions with two additional data sources. First, since 2004, the INEP agency has assessed the quality of higher education courses with an exam for first-year students and graduates called ENADE.\footnote{ENADE stands for National Students' Performance Assessment (\emph{Exame Nacional de Desempenho de Estudantes}).} I use the final scores of this exam as a measure of program quality. I also utilize data from a personal questionnaire that students answer on the ENADE exam to extrapolate the percentage of students who work while in college, the percentage of students who completed high school in public schools, and parental education levels. Second, with unique individual identifiers, I can also link students from UFBA to the Brazilian matched employer-employee dataset (RAIS), allowing me to identify whether specific students work formally while in college.\footnote{I match students by their social security number (\emph{``Cadastro de Pessoa Física"} CPF).}\fnsep\footnote{RAIS (\emph{Relação Anual de Informações Sociais}) is an annual matched employer-employee dataset, collected by the Ministry of Labor and the Ministry of Economy.} 

I use several additional datasets to measure labor market conditions. From the Demographic Censuses in 2000 and 2010, I obtain the distribution of occupations for individuals that graduated from each major. The matched employer-employee dataset allows me to compute the stock of employees and the number of new hires in each occupation, state, and time. Additionally, I use the household surveys (PNAD and PNADC) from 2002 to 2019 to estimate population counts and unemployment rates. Section \ref{sec:proxy} details how these pieces of information are used to construct the labor market measure, and there are further details in Appendix \ref{data_details}.

\begin{table}[!ht] \centering 
  \caption{Descriptive statistics}\label{desc_stat}
  \begin{adjustbox}{max width = \textwidth, width = \textwidth, center}
  \begin{threeparttable}
  \begin{tabular}{@{\extracolsep{3pt}} lccccccc} 
  \\[-1.8ex]\hline 
  \hline \\[-1.8ex]
  &\multicolumn{3}{c}{Mean}  &&\multicolumn{3}{c}{Number of Observations} \\  \cline{2-4} \cline{6-8}  \\[-.25cm]
   & Private & Public & UFBA & & Private & Public& UFBA \\[.2cm]  \hline \\[-1.8ex]
  On-time graduation&0.457&0.324&0.216&&3,129,095&1,540,563&42,154\\
  \\
  \underline{{Demographics}}\\
  \hphantom{a}Age at admission &19.310&19.246&-&&3,343,106&1,622,900&-\\
  \hphantom{a}Female & 0.609 & 0.561 &0.551&&3,343,106&1,622,900&29,550\\
  \hphantom{a}Black or Native & 0.335 & 0.404 & 0.754 && 1,992,907& 1,047,273 & 25,465 \\ 
  \\
  \underline{Program-level variables}\\
  \hphantom{a}Top-10\% programs (ENADE) & 0.051 & 0.354 & - && 3,106,390 & 1,314,787 & - \\ 
  \hphantom{a}Top-10\% programs (CPC) & 0.062 & 0.166 & - && 3,104,399 & 1,313,623 & - \\ 
  \hphantom{a}$\geq$50\% of mothers with College+ &0.147&0.254&-&&3,014,310&1,254,876 &  - \\
  \hphantom{a}$\geq$50\% from public high-schools &0.614&0.436&-&&3,014,310&1,254,876 &  - \\
  \hphantom{a}$\geq$50\% working full-time&0.310&0.080&-&&3,014,310&1,254,876 &  - \\
  \\
  \underline{Individual-level variables}\\
  \hphantom{a}Working full-time in Junior Year&-&-&0.208&&-&-&42,154\\[.1cm]
  \hline \hline \\[-1.8ex] 
  \end{tabular}
  \begin{tablenotes}
    \item \footnotesize{Notes: Descriptive statistics for the sample of students in the Higher Education Census by type of institution (private or public) and for the sample of students in UFBA. The first three columns present the sample mean and the last three columns the number of observations with non-missing information for each variable. Junior year is the year before expected graduation. Top-10\% programs classified using the ENADE score.}
\end{tablenotes}
  \end{threeparttable}
  \end{adjustbox}
\end{table} 

Table \ref{desc_stat} shows the descriptive statistics of my two main datasets. I divide the students from the Higher Education Census by institution type (public or private). In line with the discussion in section \ref{institutional_setting}, we can see significant differences across the two types of institutions. Students in private universities are more likely to graduate on time (45.7\% x 32.4\%), more likely to be women (60.9\% x 56.1\%), and less likely to be Black or Native (33.5\% x 40.4\%). I use two measures of course quality, one based on the ENADE exam, computed as the average of students graduating from a given program. The other (CPC) is produced by the Ministry of Education, combining data from the ENADE exam and information on the program infrastructure and faculty composition. While 35.4\% of students in public institutions are enrolled in programs classified in the top-10\% of ENADE score, only 5.1\% of students in private schools are in these top programs. We have a similar picture for the CPC measure, but with a lower fraction of students in public universities enrolled in the top 10\% programs. In terms of socioeconomic status, only 14.7\% (24.4\%) of students in private (public) schools are enrolled in programs in which at least 50\% of students have mothers with a college degree or more, and 61.4\% are enrolled in programs where more than half of students graduated in public high schools. 
Strikingly, 31\% of students in private universities are in programs where more than half of students work full-time while in college, versus just 8\% in public institutions. Students at UFBA have lower on-time graduation rates (21.6\%) and a much higher proportion of Black students (75.4\%) --- this is in line with the overall demographics of the state, which is the state with the highest proportion of Black people in the country. For the UFBA sample, 1 in 5 students works full-time the year before their expected graduation. 

\section{Empirical Strategy}

The empirical strategy explores variation in employment conditions over time, geography, and chosen majors. Student $i$ was admitted to college in year $t_0(i)$, in major $m(i)$, and, given the length of their major, they are expected to graduate in year $t_1(i)$. They are studying in state $s(i)$ in program $p(i)$, a combination of major-institution-schedule. My main specification regresses the dummy outcome variable for on-time graduation $Y_i$ on an employment measure that varies across time-space-major ($H_{t_1(i),s(i),m(i)}$). 
\begin{equation}
Y_i = \beta H_{t_1(i),s(i),m(i)} + \eta_{t_0(i),m(i)} + \nu_{p(i)} + \gamma X_i + \varepsilon_i \label{eq:main_specification}
\end{equation}

I control flexibly for the time of entry by adding admission-year fixed-effects for each major ($\eta_{t_0(i),m(i)}$). Tertiary enrollment in Brazil rose steadily in the first 15 years of the 20th century. Therefore, I am careful in my analysis to avoid capturing the effects of this unrelated trend. The introduction of the $\eta_{t_0(i),m(i)}$ term still allows for some variation in the time of graduation because programs have different duration. I also add program fixed effects ($\nu_{p(i)}$) that absorb fixed unobserved factors that vary by major-institution-schedule of study, and individual level controls ($X_i$) that include demographic controls for gender, race, and age at admission. In section \ref{institutional_setting}, I discussed the differences between public and private institutes in Brazil and why we expect different results along this dimension. Therefore, I analyze public and private institutions separately. I clustered the standard errors at both the major and the state level (\emph{two-way cluster}), allowing for arbitrary correlation within states as well as for students in the same major.\footnote{When showing the results for the UFBA sample, I clustered the standard errors at the major level because, for these exercises, all students graduate from the same state.}

I am interested in analyzing $\beta$, which will capture the effects of the current labor market conditions on on-time graduation for students in public and private universities. The identification assumption is that, conditional on the included covariates, there are no unobserved components correlated with the labor market measure and the graduation decision. While this assumption is not testable, I show in the Appendix table \ref{tab:balance} that, conditioning on the set of fixed effects (excluding the demographic cells fixed effects), individual characteristics of the students (gender, race, and age of entry) are not correlated with my labor market measure. One potential threat is, for instance, if labor market shocks are correlated over time, and negative shocks affect their performance in the first years of college. If that is the case, then what I interpret as a delaying effect is, instead, just students being unable to graduate because they did not accumulate enough credits to graduate on time. Other concerns are the selectivity of state and major of graduation, which could be correlated with future labor market shocks. I discuss how my results are robust to these concerns in Section \ref{sec:robustness}.

\subsection{Employment Measure}\label{sec:proxy}

The exercise requires a measure for the labor market conditions faced by student $i$, majoring in $m(i)$, who is expected to graduate at $t_1(i)$, in state $s(i)$. Most of the literature covered by \cite{vonwachter2020} uses the unemployment rate at some sub-national level (state, provinces). I leverage the matched employer-employee dataset to have finer variation at the major level, similar to the major-specific unemployment rates by \cite{altonji2016}. This allows me to explore variations across time, geography, and majors. 

I obtain the distribution of occupations across majors for all individuals with a college degree from the Demographic Census in 2000 and 2010. I construct the vector $w_m = (w_m^1,w_m^2,\hdots,w_m^{N_{occ}})$ for each major $m$, where each component is the proportion of individuals from major $m$ that are employed in each of the $N_{occ}$ occupations. These weights vary greatly across majors and are fairly constant over time.\footnote{I compute the correlation between $w_m$ and $w_{m^{\prime}}$ for all pairs of majors --- the median value is 0.15, evidencing that they vary across majors. I also show that weights are constant over time by computing the weights separately for 2000 and 2010 and computing the correlation of $w_{m,2000}$ with $w_{m,2010}$ for all majors --- the median correlation is 0.87. Figure \ref{fig:weights_assessment} shows the full distribution of these two metrics.}

Using the matched employer-employee data, I compute for each state ($s$) and occupation ($o$) the number of new hires in a given semester ($t$), defined as $h_{ost}$.\footnote{To avoid seasonal effects, I aggregate this measure annually. For the 1st semester of year $t$, I aggregate hiring from the first semester of $t$ with the second semester of $t-1$. For the second semester of year $t$, I use hires from both semesters of $t$.} Occupations are defined in the 4-digit level, using a crosswalk between the codes in the Demographic Censuses and RAIS. Importantly, to avoid mechanical effects, I compute new hires only for individuals aged 27 or above.\footnote{{This restriction eliminates a direct link between the number of hires and the supply of graduates of the same age. However, the new hires for the population above 27 years could still be affected by the supply of new graduates. This indirect link would underestimate my effects since it generates a positive correlation between student postponement and new hires (above 27). I expected business cycle-induced movements to dominate the time-spatial-major variation of new hires. If I instead use a lagged version of the MWH variable, which does not suffer from mechanical effects, I obtain very similar results.}}\label{pag:MWH_mechanical} Given my age of entry restrictions (17-22), less than 0.01\% of my sample is 27 at expected graduation, and more than 96\% are younger than 27 even two years after expected graduation.

With $w_m$ and $h_{ost}$, I construct the Major-Weighted Hiring measure as a weighted sum of the new hires for each occupation and state: 
\begin{equation}
 \text{MWH}_{tsm} = \sum_o w_m^o h_{ost}
\end{equation}
Therefore, students expected to graduate in majors whose typical occupations are hiring more will face a larger $MWH$ than students graduating from majors whose typical occupations have a hiring freeze. I will use $\log(MWH_{tsm})$ as the labor market measure in equation \ref{eq:main_specification} to capture \emph{percentual} variations of the labor market conditions. Therefore $\beta$ can be directly read as the effect of a 100\% variation on the (weighted) hires. For the UFBA sample, I compute the major-weighted hiring with a similar procedure.

In order to compute reliable estimates for $MWH_{tsm}$, I apply sample restrictions to remove occupations with small coverage and changing codes over time. These restrictions are detailed in the Appendix \ref{data_details}. Reassuringly, I conduct a sensitivity analysis in the Appendix \ref{app:sensitivity} showing that the results are not sensitive to these restrictions.

\begin{figure}[!ht]
 \centering
 \caption{Economic activity measures}
 \label{fig:emp_measures}
 \includegraphics[width=\textwidth]{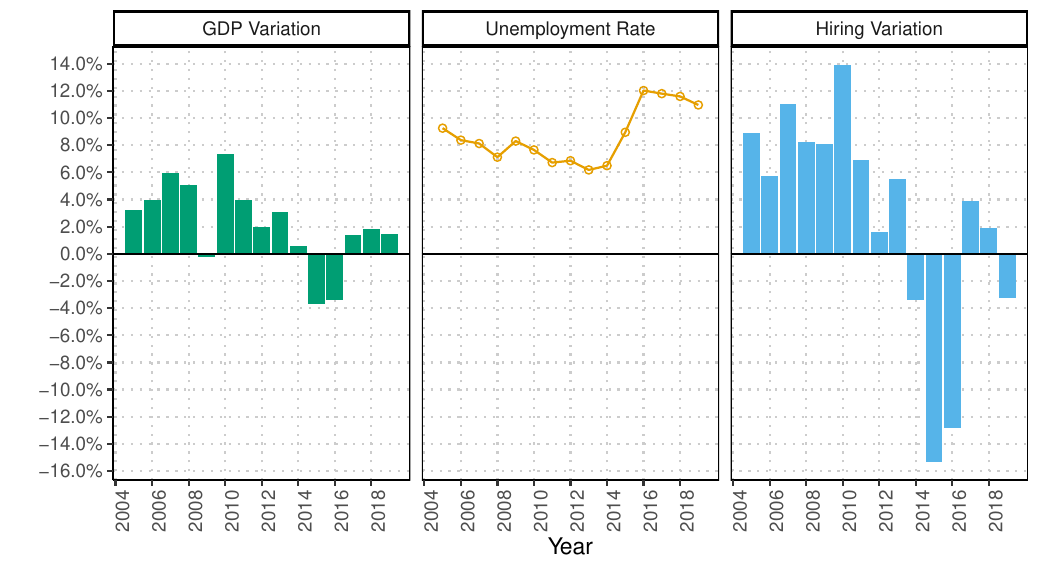}
 \caption*{\footnotesize{Notes: The first panel shows the annual variation of the Brazilian GDP. The second panel shows the national overall unemployment rate across time. The third panel shows the annual variation of the hiring measure.}}
\end{figure}

\begin{figure}[!ht]
 \centering
 \caption{MWH for major-state pairs across time}
 \label{fig:mse_variation}
 \includegraphics[width=\textwidth]{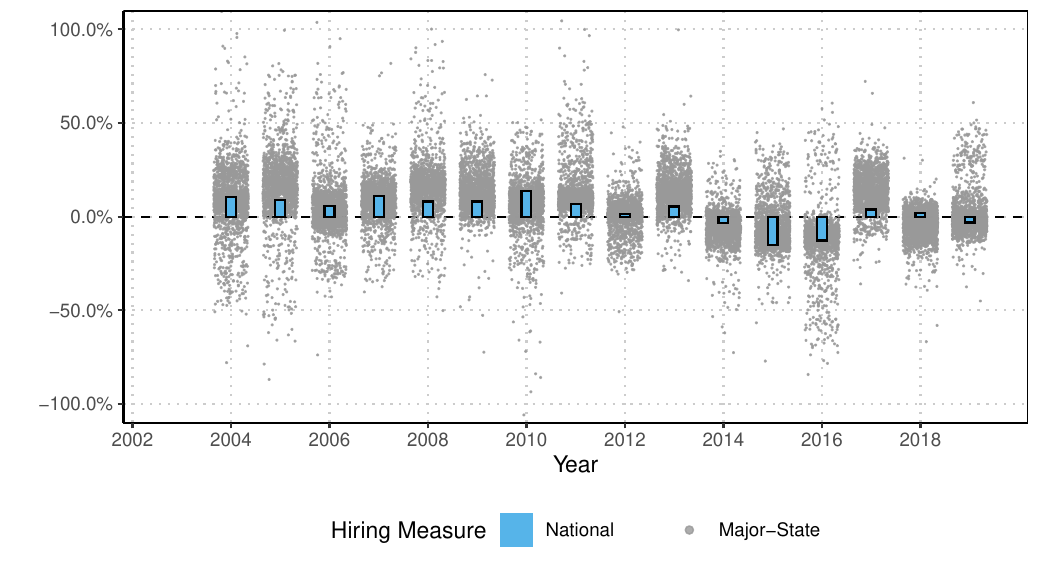}
 \caption*{\footnotesize{Notes: The blue bars are the annual national variation for the hiring variable. Each gray dot shows the annual variation (first differences) of the major-weighted hiring measure for one major-state pair. I spread the dots on the x-axis to improve the visualization --- all dots in the same ``block" are from the same year.}}
\end{figure}

Figure \ref{fig:emp_measures} shows that the proposed measure closely tracks changes in the unemployment rate and variations in the GDP. This figure uses the variation in the annual hiring counts for the entire country. Brazil experienced a major recession in 2014--2016, during which the GDP fell by 4\% for two consecutive years, increasing the unemployment rate by 5.5 percentage points from 6.5\% to 12\%. Figure \ref{fig:mse_variation} plots the variation of the hiring measure for all major-state pairs across time (each gray dot), which is the variation my analysis explores. Brazil has 26 states and 1 federal district, and I have information for 64 majors, yielding 1,728 major-state pairs. Besides tracking the variations in the GDP and employment closely, MWH has two main advantages over these measures: it has a finer variation at the major level and is likely close to the real consideration set of students. A student graduating in engineering responds more to how occupations typically hiring engineers, like industry, construction, and the financial sector, are trending than in health-related occupations. 

\section{Results}

For ease of exposition and to better align with existing literature, I multiply the coefficient of interest ($\beta$, from equation \ref{eq:main_specification}) by minus one. Therefore all the coefficients can be interpreted as a 100\% decrease in the major-weighted hiring measure. 

\subsection{Main Results}

Table \ref{res:main_results} presents the baseline results. The first column shows that by decreasing the hiring measure by 1\%, on-time graduation is reduced by 0.014 percentage points. In the second column, I estimate this effect separately for public and private universities. We can see that a 1\% reduction in hiring implies that students in public universities are 0.08pp less likely to graduate. For students in private institutions, we see a small positive estimate. I reject that the two effects are the same with a p-value of 0.032. The specification in the second column includes fixed effects for each major-state and a quadratic trend on admission year. In the third column, I add a richer set of program fixed effects defined at the major-institution-schedule level. In the fourth column, I include demographic fixed effects defined by gender, race, and age at admission. 

\begin{table}[!ht] \centering 
    \begin{adjustbox}{max width = \textwidth, width = \textwidth, center}
    \begin{threeparttable}
    \caption{Main results}\label{res:main_results}
    \begin{tabular}{lcccccc} 
    \\[-1.8ex]\hline 
    \hline \\[-1.8ex]
    \multicolumn{1}{r}{\emph{Outcome:}} & \multicolumn{6}{c}{\emph{On-time graduation}}\\
    \addlinespace
    & (1) & (2) & (3) & (4) & (5) & (6) \\[.8ex] \hline \\[-.8ex]
    \addlinespace
    Hiring & \num{-0.014} & & & & & \\
     & (\num{0.024}) & & & & & \\
     & {}[\num{0.577}] & & & & & \\
     \\
    Hiring x Public & & \num{-0.081} & \num{-0.072} & \num{-0.071} & \num{-0.076} & \num{-0.070}\\
     & & (\num{0.031}) & (\num{0.031}) & (\num{0.030}) & (\num{0.033}) & (\num{0.031})\\
     & & {}[\num{0.015}] & {}[\num{0.027}] & {}[\num{0.027}] & {}[\num{0.028}] & {}[\num{0.032}]\\
     \\
    Hiring x Private & & \num{0.004} & \num{0.014} & \num{0.019} & \num{-0.023} & \num{-0.003}\\
     & & (\num{0.033}) & (\num{0.027}) & (\num{0.028}) & (\num{0.037}) & (\num{0.040})\\
     & & {}[\num{0.898}] & {}[\num{0.608}] & {}[\num{0.505}] & {}[\num{0.540}] & {}[\num{0.934}]\\
    \midrule
    N Obs & {4,058,758} & {4,058,758} & {4,058,758} & {4,058,758} & {4,058,758} & {4,058,758}\\
    \\
    p-value ($\beta_{\text{public}}=\beta_{\text{private}}$) & - & \{0.032\} & \{0.012\} & \{0.015\} & \{0.132\} & \{0.084\} \\ 
    \\
    Major-State FE &$\checkmark$ & $\checkmark$ &- & - & - & - \\
    Program FE & -&-& $\checkmark$& $\checkmark$& $\checkmark$ & $\checkmark$\\
    Demographics & -& - & $\checkmark$& $\checkmark$&$\checkmark$ & $\checkmark$\\
    \addlinespace
    \multirow{2}{*}{Time Trend} & \multirow{2}{*}{Quadratic} & \multirow{2}{*}{Quadratic} & \multirow{2}{*}{Quadratic} & \multirow{2}{*}{Quadratic} & Admission & Major-Admission \\
    &&&&&Time FE & Time FE\\
    \hline 
    \hline \\[-1.8ex] 
    \end{tabular} 
    \begin{tablenotes}
    \item \footnotesize{Notes: The table presents the estimation of $\beta$ from equation \ref{res:main_results}. I multiply the coefficient by minus one, therefore, we can see the coefficients as a decrease of 100\% of the weighted hiring measure. The recession between 2014-2016 reduced the weighted hiring measure by 30\%. The first column presents the overall results. In columns 2-6, I interact all variables with an indicator of whether the students belong to a private or public institution. Columns 2-6 differ in the set of control variables included in each specification. Standard errors are clustered at both the major and state levels (\emph{two-way clustering}). The p-value of the test whether the effect for public and private institutions are the same is provided for each specification. FE stands for fixed effects.}
    \end{tablenotes}
    \end{threeparttable}
    \end{adjustbox}
\end{table}

Section \ref{institutional_setting} showed how enrollment in college education in Brazil rose in the first 15 years of the 2000s. I want to be careful not to capture effects that can stem from the rise in enrollment. The first four columns from Table \ref{res:main_results} include a quadratic trend on admission year, which act as a parsimonious control for any time-varying trend in tertiary education that may relate to the time of graduation. In the fifth column, I replace the quadratic trend with a fully flexible admission time fixed effects, which does not impose any parametric restriction on the time trends. We can see that the results continue to be really close. In the sixth column, I allow the admission time fixed effects to be major-specific. It can therefore account for any major-specific trends.

I will use the specification in the last column as the benchmark since it flexibly controls for time trends, allowing for major-specific trends. Reducing hiring by 1\% implies a reduction of on-time graduation by 0.07pp for students in public universities. In the 2014--2016 recession, the average reduction in the hiring measure was 30\%, implying a reduction of 2.1pp on on-time graduation. Considering the average on-time graduation rate in public universities of 32.4\%, the effect translates to a 6.5\% reduction in the on-time graduation rate. We can see that across all specifications in Table \ref{res:main_results}, the estimates for students in public universities are robust, ranging from 0.070--0.081. In contrast, the estimates for students in private institutions are small, not statistically significant, flipping signs. To decide the graduation time, students in private universities are likely in corner solutions when comparing the costs and benefits discussed in section \ref{sec:theoretical_framework}. 

Public and private universities have several different characteristics: public universities are, on average, more selective, with better quality courses, with fewer students working while in college compared to their private counterparts. Importantly, public universities are tuition-free. In the Appendix \ref{app:reweigthing}, I reweight the observations from private institutions to have the same distribution of the public universities along the following dimensions: majors offered, demographics of students, the proportion of students from public high schools, and quality. In none of the reweighting exercises the public and private estimates are close. Reweighting for quality produces the estimates where the public-private gap is the smallest. Nevertheless, the point estimate for private institutions is only one-quarter of the estimate for public universities. 

While I do not have information on tuition, I speculate that the fact that public universities do not charge tuition may be an important factor explaining these results. While students in private universities would need to pay tuition for the extra semesters of postponement, this direct cost is zero for students in public universities. This explanation is aligned with the findings of \cite{garibaldi2012}, who finds that increasing tuition reduces late graduation in Italy. The result also is consistent with the cross-country evidence from \cite{brunello2003}, which shows a shorter college duration for students facing higher tuition costs. 

{The estimates above benefit from variation in time, geography, and major, using administrative data on the universe of hires in Brazil. In Appendix Table \ref{res:robustness_unemp}, I reproduce the same results from Table \ref{res:main_results}, using as labor market measure the state unemployment rate for college graduates, a common measure used in the graduating-in-a-recession literature. The estimates are noisier but display the same qualitative patterns. The delaying effects for students in public universities range from a reduction of 0.21--0.48pp for each increase in 1pp of the unemployment rate. The estimates for students in private universities flip signs and are very small in the preferred specifications.}\label{pag:page_robustness_unemp}

\begin{figure}[!ht]
 \centering
 \caption{Effects on \emph{cumulative} graduation by semester relative to expected graduation}
 \label{fig:panel_effects}
 \includegraphics[width=.8\textwidth]{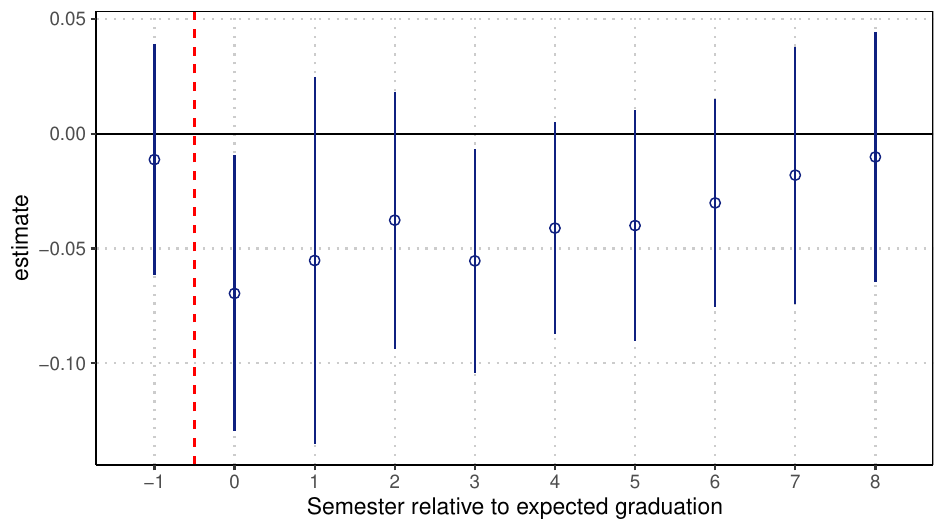}
 \caption*{\footnotesize{Notes: The figure plots the estimation of $\beta$ from equation \ref{res:main_results} for students in public universities. I multiply the coefficient by minus one. Therefore, we can see the coefficients as a decrease of 100\% of the weighted hiring measure. The recession between 2014--2016 reduced the weighted hiring measure by 30\%. Each dot is estimated separately, using as the outcome variable whether the student has already graduated in semester $\tau$ relative to expected graduation (cumulative graduation measure). The dots are the estimates, and the lines represent the 95\% confidence intervals. Standard errors are clustered at the major and state levels (\emph{two-way clustering}). All regressions include fixed effects for program, time of admission, fall semester, and demographic cells (gender, race, and age).}}
\end{figure}

The previous results showed that fewer public university students graduate on time when facing a more depressed labor market. However, we still need to determine whether they postponed graduation or did not graduate. Figure \ref{fig:panel_effects} plots the effects of the baseline regression estimated separately for each semester relative to expected graduation for students in public institutions. The outcome $Y^{\tau}_i$ is now equal to 1 if student $i$ graduated up to semester $\tau$, that is, the \emph{cumulative} graduation up to $\tau$ semesters after expected graduation. The estimate at (relative) semester 0 is the same as in Table \ref{res:main_results}. 

One semester before expected graduation ($\tau=-1$ in the graph, on the x-axis), the point estimate is close to zero, dropping to -0.070 in the semester of expected graduation. After one year, the effect is around -0.037 and quickly becomes indistinguishable from zero. The fact that it does not have a significant effect four years after expected graduation shows that the labor market conditions can affect graduation timing but not whether students ultimately graduate. Note that for all regressions, $MWH$ is fixed at the hiring rate observed at the expected graduation semester. It is worth emphasizing that I am restricting to students enrolled in their expected graduation year.

Using the estimated effects displayed in Figure \ref{fig:panel_effects}, I can compute the average delay response to changes in the hiring rates. The 30\% decrease in hiring observed in the 2014--2016 regression implies a 0.112 increase in average college duration, 1 out of 9 students delaying by one semester, or 1 out of 18 students delaying graduation by one year. In the Appendix \ref{app:tobit}, I also obtain this number directly using a Tobit regression, where the outcome is the censored semester of actual graduation. 

\subsection{Heterogeneity}\label{sec:heterogeneities}

\begin{figure}[!h]
 \centering
 \caption{Heterogeneity}\label{fig:res_heterogeneity}
 \includegraphics[width=\textwidth]{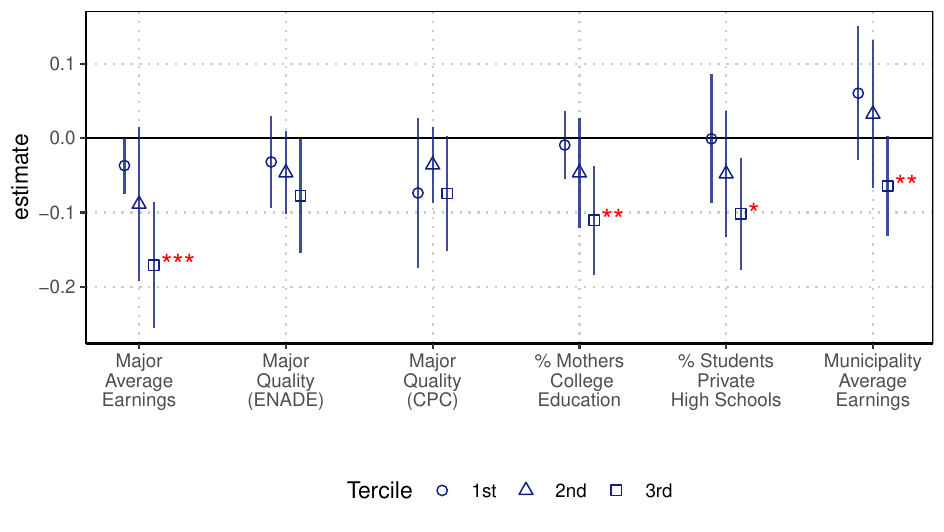}
 \caption*{\footnotesize{Notes: The figure plots the estimation of $\beta$ from equation \ref{res:main_results} for students in public universities interacted with terciles for each variable. I multiply the coefficient by minus one, therefore, we can see the coefficients as a decrease of 100\% of the weighted hiring measure. The recession between 2014--2016 reduced the weighted hiring measure by 30\%. Each group of heterogeneity is estimated separately. The circles are the effects for the 1st tercile, the triangles for the 2nd, and the squares for the 3rd tercile. The circles, triangles, and squares are the estimates, and the lines represent the 95\% confidence intervals. Standard errors are clustered at the major and state levels (\emph{two-way clustering}). All regressions include fixed effects for program, time of admission, fall semester, and demographic cells (gender, race, and age). The red stars represent the p-values of the tests for whether the effects for the second and third terciles are the same as the one for the first tercile. 3 stars are used for p-values inferior a 1\%, 2 starts for 5\%, and 1 star for 10\%.}}
\end{figure}

In this section, I investigate heterogeneities in the effects for students in public universities. Figure \ref{fig:res_heterogeneity} presents the results. The first shows the heterogeneous effects according to the average major earnings tercile, computed using the total earnings for all individuals in a given major from the Demographics Censuses in 2000 and 2010, net from gender, race, and age effects.\footnote{I use the total earnings of individuals between 25 and 65 years that graduated for each major, pooling data from the Demographic Census in 2000 and 2010. I first residualized earnings on gender, race, and age fixed effects.} Students in majors with higher predicted earnings are more likely to delay graduation in response to labor market shocks. The two red stars close to the coefficient for the third tercile show that we can reject, at 5\% level, that $\beta_{\text{3rd Tercile}} = \beta_{\text{1st Tercile}}$.

I compute heterogeneity across two quality measures, the first using the average scores of the graduating students in the ENADE exam and the second a quality measure computed by the Ministry of Education (CPC). Using the average score of graduating students at ENADE, we see weak evidence that students in better programs are more likely to delay in response to a labor market shock. We do not see the same ordering across the three terciles using the CPC measure. 

The last three results turn to the socioeconomic conditions and proxies for family resources. Students in majors with a higher proportion of mothers with college education and a higher proportion of students from private high schools exhibit larger effects. In both cases, we reject that the effects for the third tercile are the same as for the first tercile. The last result splits students by the average municipality earnings, computed with the same procedure as the major average earnings. Students born in richer municipalities exhibit larger effects. 

\begin{table}[!ht] \centering 
    \begin{adjustbox}{max width = \textwidth, width = .75\textwidth, center}
    \begin{threeparttable}
    \caption{Heterogeneity by gender and race}\label{res:het_gender_race}
    \begin{tabular}{lccc} 
    \\[-1.8ex]\hline 
    \hline \\[-1.8ex]
    \multicolumn{1}{r}{\emph{Outcome:}} & \multicolumn{2}{c}{\emph{On-time graduation}}\\
    \addlinespace
    & (1) & (2) & p-value \\
    \addlinespace \\[.8ex] \hline \\[-.8ex]
    &Men & Women & $(\beta_{\text{Men}}=\beta_{\text{Women}})$\\ \cmidrule(l{3pt}r{3pt}){2-2}\cmidrule(l{3pt}r{3pt}){3-3} \cmidrule(l{3pt}r{3pt}){4-4}
    \addlinespace
    Effect & \num{-0.080} & \num{-0.074} & p-value \\
     (s.e.) & (\num{0.032}) & (\num{0.033})&\{0.555\} \\
     
     [p-value]& [\num{0.020}] & [\num{0.037}]\\
    \\[.3cm]
    &White/Asian & Black/Native & $(\beta_{\text{White/Asian}}=\beta_{\text{Black/Native}})$\\ \cmidrule(l{3pt}r{3pt}){2-2}\cmidrule(l{3pt}r{3pt}){3-3}\cmidrule(l{3pt}r{3pt}){4-4}
    \addlinespace
     Effect & \num{-0.057} & \num{-0.047} & p-value \\
     (s.e.) & (\num{0.041}) & (\num{0.029})&\{0.753\}\\
     
     [p-value]& [\num{0.179}] & [\num{0.115}]\\
    \\
    
    \hline 
    \hline \\[-1.8ex] 
    \end{tabular} 
    \begin{tablenotes}
    \item \footnotesize{Notes: The table presents the estimation of $\beta$ from equation \ref{res:main_results} for students in public universities interacted with gender and race. I multiply the coefficient by minus one, therefore, we can see the coefficients as a decrease of 100\% of the weighted hiring measure. The recession between 2014-2016 reduced the weighted hiring measure by 30\%. Each group of heterogeneity is estimated separately. Standard errors are clustered at both the major and state levels (\emph{two-way clustering}). All regressions include fixed effects for program, major-admission time, fall semester, and demographic cells (gender, race, and age). The third column shows in brackets the p-value of the test whether the effect for men is the same for women (first row) and the effect for White/Asian is the same for Black/Native students (second row).}
    \end{tablenotes}
    \end{threeparttable}
    \end{adjustbox}
\end{table}

Table \ref{res:het_gender_race} presents heterogeneities at the student level, by gender and race. We do not reject the null hypothesis that the effect for men and women are the same. We have a slightly smaller estimate for Black and Native students, which would be consistent with the heterogeneity results at the program level. However, we do not reject that those coefficients are the same. Moreover, the exercise by race needs to be analyzed with care due to the high level of missingness for this variable (between 35-40\%). 

\subsection{Robustness}\label{sec:robustness}

\begin{figure}[!ht]
 \centering
 \caption{Robustness}
 \label{fig:robustness}
 \includegraphics[width=\textwidth]{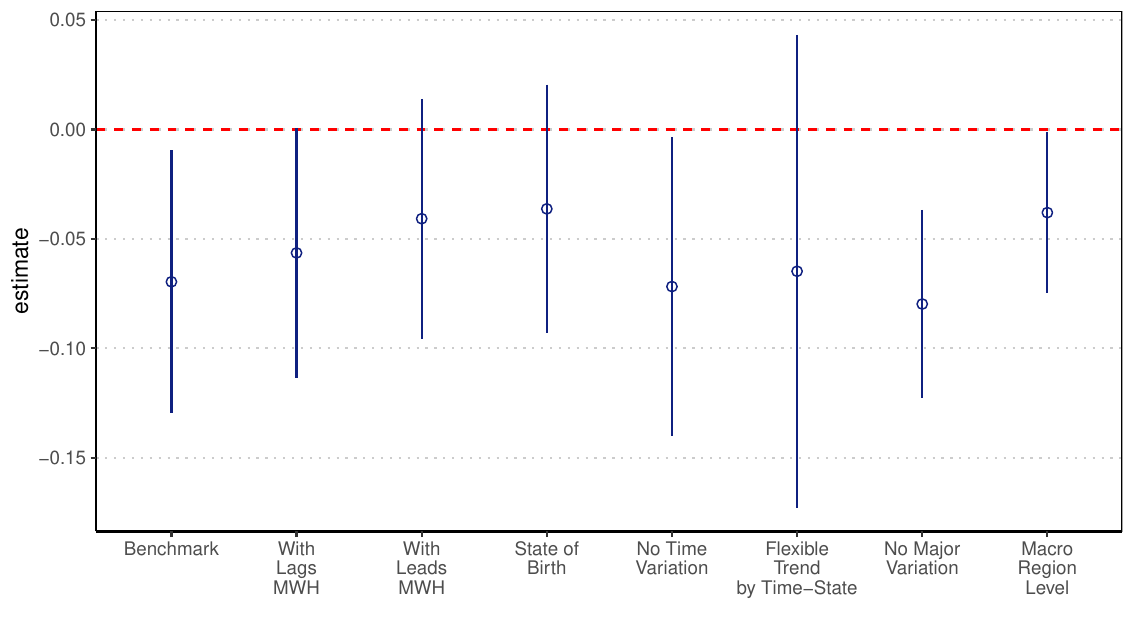}
 \caption*{\footnotesize{Notes: The figure plots the estimates of $\beta$ from equation \ref{res:main_results} for students in public universities. I multiply the coefficient by minus one, therefore, we can see the coefficients as a decrease of 100\% of the weighted hiring measure. The recession between 2014--2016 reduced the weighted hiring measure by 30\%. The circles are the point estimates, and the lines represent the 95\% confidence interval. Standard errors are clustered at the major and state levels (\emph{two-way clustering}). Each circle is a different robustness exercise detailed in the main text.}}
\end{figure}

In this section, I explore other threats to my identification strategy and the robustness of the results. Figure \ref{fig:robustness} presents the results. The first dot shows the point estimate of the benchmark result with a line segment representing the 95\% confidence interval. 

One concern with my strategy is that employment conditions are correlated over time. Facing a worse labor market near expected graduation could also imply that the student also faced worse labor market conditions during their college studies. Therefore, what I interpret as students avoiding entering a depressed labor market could instead be students not being able to graduate because of bad shocks in the labor market interacting with their proficiency in the program, affecting their failure rate and credit accumulation. While this alternative mechanism could be plausible, it is not consistent with the heterogeneities I found in Section \ref{sec:heterogeneities} --- the interactions of bad labor market shocks with student performance would imply larger effects for students with lower socioeconomic status, while I found the exact opposite. Nevertheless, I augment specification \ref{eq:main_specification}, including lagged hiring variable for up to 3 years, spanning the entire period most of the students are in college in my sample. The result is the second estimate in Figure \ref{fig:robustness}, showing a very similar effect. In a similar concern, it could be that students are responding to future labor market conditions and not the conditions at graduation, as I interpret them. The third result includes leads of the variable of interest for the next three years, yielding a similar result. 

The strategy employed in this paper explores spatial variation on employment using the state where students are expected to graduate, as is commonly employed in the existing literature. However, it could be the case that students sort into colleges depending on the labor market conditions, which would violate my identification strategy. In the fourth result, I replace the MWH for the state of birth of students and not the state where the institution is, obtaining a slightly smaller estimate. 

The main specification includes major-admission time fixed effects, to control for the trends in higher education enrollment in Brazil. The fifth result shows that I obtain similar results even if I completely eliminate any use of time variation by including graduating semester fixed effects. The following result also relaxes the trend assumption by further interacting the major-admission time fixed effects with states. There is a significant loss of precision since comparisons across states are eliminated. 

In the eighth result in figure \ref{fig:robustness}, I use the hiring measure computed without the major-specific weights, obtaining a similar result.\footnote{I compute average hirings for all employees with college degrees for each state.} The last result computes hiring at the macro-region level instead of states. Brazil has 135 macro-regions, which subdivide the 27 states. This approach has the advantage of being more local and bringing more variation to the estimation. However, it can also bring more measurement error since the labor market in consideration can span different macro-regions for several students. 

{The results show how students in public universities respond to local labor market conditions adjusting the graduation timing, but not whether to graduate. In contrast, students in private schools do not respond to market conditions. Students could also respond to labor shocks by starting graduate programs. Unfortunately, the data does not allow me to investigate this margin. However, I do not believe it affects my conclusions since only 6--7\% of individuals with college degrees in Brazil also have a graduate degree. In addition, a significant fraction of the graduate school enrollment is not immediately after college graduation. Nevertheless, my results should be interpreted as a net effect from any potential increase in graduation due to individuals adjusting graduation timing to start graduate programs.}\label{pag:grad_outcome}

In addition to these robustness exercises, I conduct four extra exercises. First, as a placebo check, I run the main specification \ref{eq:main_specification} using lagged values of the weighted hiring measures. The results are presented in the Appendix figure \ref{fig:placebo}. Using the 1-year lagged hiring measure, I still estimate a significant result, half of the magnitude of our benchmark estimate. Reassuringly, the estimates with more than two-year lags are fairly small and indistinguishable from zero. Second, the causal interpretation relies on the identification assumption that, conditional on our rich set of fixed effects, there are no unobserved factors correlated with the employment measure and on-time graduation. While this assumption is untestable, I show in the Appendix table \ref{tab:balance} that the demographics of students are not correlated with the employment measure. There is also a concern that the relationship between employment and on-time graduation is not linear. In Appendix Figure \ref{fig:np}, I first residualized both the outcome (on-time graduation dummy) and hiring measure on all the other variables in equation \ref{eq:main_specification}. I present ten bins with average residualized hiring measure and average residualized on-time graduation. We can see that all bins closely resemble the linear estimate. Lastly, I also show in the Appendix \ref{app:sensitivity} that the results are robust to changes in the sampling definitions and to excluding years where the data was linked over time using the matching algorithm.

\subsection{Course credits and working while in college}

While the Higher Education Census provides broad coverage of all students in higher education in Brazil, it does not include information about their course load towards completing the degree and detailed individuals level information. I overcome these two issues by accessing detailed data from one large public university in the state of Bahia --- the federal university of the state (flagship state university) UFBA. Unfortunately, this data comes from one state, so the hiring measure can only vary by time and major. 

First, Table \ref{res:ufba_mainresults} shows the main results for this sample. In the first column, a 1\% reduction in the major-weighted hiring rate decreases the on-time graduation rate by 0.18 percentage points. This is larger than our estimates for public universities in the main sample. The last three columns of Table \ref{res:ufba_mainresults} show the heterogeneity by admission score terciles, computed within each major-year. We can see that students at the top of the distribution exhibit larger effects. 

\begin{table}[!ht]
      \caption{UFBA --- Main results}\label{res:ufba_mainresults}
     \begin{adjustbox}{max width = \textwidth, width = .6\textwidth, center}
     \begin{threeparttable}
     \begin{tabular}[t]{lccccc}
      \hline \hline 
      \multicolumn{1}{r}{\emph{Outcome:}} & \multicolumn{5}{c}{\emph{On-time graduation}}\\\addlinespace
      & Overall &\hphantom{a}& \multicolumn{3}{c}{by Admission Score Tercile} \\ \cmidrule(l{3pt}r{3pt}){4-6}
      & && 1st & 2nd & 3rd
      \\[.8ex] \hline \\[-.8ex]
      \addlinespace
      Hiring & \num{-0.181} && \num{-0.117}& \num{-0.216}& \num{-0.220}\\
      
      (s.e.) & (\num{0.046}) && (\num{0.056})& (\num{0.067}) & (\num{0.063})\\
      
      [p-value]& {}[\num{0.000}] && {}[\num{0.040}]& {}[\num{0.002}]& {}[\num{0.001}]\\
      \\
      N obs & {41,475} && \multicolumn{3}{c}{34,477}\\
     \hline \hline 
      \end{tabular}
     \begin{tablenotes}
     \item \footnotesize{Notes: The table presents the estimation of $\beta$ from equation \ref{res:main_results} for the students at UFBA. I multiply the coefficient by minus one, therefore, we can see the coefficients as a decrease of 100\% of the weighted hiring measure. The recession between 2014-2016 reduced the weighted hiring measure by 30\%. The first column presents the overall result. In columns 2-4, I interact all variables with an indicator for the student admission score tercile (computed for each within program-year). Standard errors are clustered at the major level. All regressions include fixed effects for program, major-admission time, fall semester, and demographic cells (gender, race, and age).}
     \end{tablenotes}
     \end{threeparttable}
     \end{adjustbox}
\end{table}

I also use UFBA data to investigate how credit accumulation in the program is affected by the local labor market conditions. Table~\ref{res:ufba_credits_before} uses our main specification, with the outcome variable being the number of credits accumulated in different semesters before expected graduation, measured in courses-equivalent.\footnote{The average course has 78 credits.} We can see that the employment measure is not associated with lower credit accumulation in the first years of the program. In the semesters immediately before expected graduation, cohorts that experienced a more depressed labor market obtained fewer credits. They may indicate the mechanism through which students delay graduation: by taking fewer courses in the semesters near expected graduation. 

\begin{table}[ht!]
    \caption{UFBA --- Credits obtained before graduation }\label{res:ufba_credits_before}
\begin{adjustbox}{max width = \textwidth, width = .9\textwidth, center}
\begin{threeparttable}
\begin{tabular}[t]{lccccccc}
      \hline \hline \\[-.2cm]
      \multicolumn{1}{r}{\emph{Outcome:}} & \multicolumn{7}{c}{\emph{Credits Obtained}}\\
      \addlinespace 
Semesters before expected graduation:& 7& 6& 5 & 4 & 3 & 2 & 1 \\  
\midrule
Hiring & \num{-0.001} & \num{-0.015} & \num{1.111} & \num{0.140} & \num{-0.191} & \num{-1.930} & \num{-1.816}\\
 (s.e.)& (\num{0.001}) & (\num{0.011}) & (\num{0.858}) & (\num{0.216}) & (\num{0.524}) & (\num{0.758}) & (\num{1.017})\\

 [p-value] & {}[\num{0.607}] & {}[\num{0.195}] & {}[\num{0.199}] & {}[\num{0.518}] & {}[\num{0.716}] & {}[\num{0.013}] & {}[\num{0.077}]\\
\\
N Obs & \num{32167} & \num{35670} & \num{38527} & \num{41221} & \num{42806} & \num{43800} & \num{43800}\\
\hline \hline 
      \end{tabular}
\begin{tablenotes}
\item \footnotesize{Notes: The table presents the estimation of $\beta$ from equation \ref{res:main_results} for students in UFBA. I multiply the coefficient by minus one, therefore we can see the coefficients as a decrease of 100\% of the weighted hiring measure. The recession between 2014-2016 reduced the weighted hiring measure by 30\%. In each column the outcome variable is the total number of credits obtained in semester $\tau$ relative to expected graduation, where $\tau\in[1,7]$. I divide the number of credits by 78 to be read as the number of equivalent courses. Standard errors are clustered at the major level. All regressions include fixed effects for program, time of admission, fall semester, and demographic cells (gender, race and age).}
\end{tablenotes}
\end{threeparttable}
\end{adjustbox}
\end{table}

Lastly, I also perform heterogeneity analyses at the individual level. I start with whether the student was working formally in the year before expected graduation. We can see in the first row of Table \ref{res:ufba_het} that the effect is close to zero for students already working formally in the year before expected graduation. In contrast, those that were not working decreased their likelihood of on-time graduation by 0.228pp when facing a 1\% decrease in the hiring rate. In the next three sets of results, I analyze proxies for socioeconomic status at the individual level. They all point to the same conclusion: larger effects for the students with better socioeconomic status and more family income. 

\begin{table}[!ht]
    \caption{UFBA --- Heterogeneity }\label{res:ufba_het}
\begin{adjustbox}{max width = \textwidth, width = .75\textwidth, center}
\begin{threeparttable}
\begin{tabular}[t]{lcccc}
      \hline \hline  \addlinespace
      & Hiring & Std Error & P-value & P-value Difference\\
      \hline \addlinespace
 \textbf{Working in Junior Year} \\ 
\hphantom{a} Yes & \num{-0.010}  & (\num{0.081})  & {}[\num{0.907}] & - \\
\hphantom{a} No & \num{-0.228}  & (\num{0.044}) & {}[\num{0.000}] & \{0.003\}\\
\\
\textbf{Public High School}\\
\hphantom{a} Yes & \num{-0.120}  & (\num{0.053})  & {}[\num{0.026}]& - \\
\hphantom{a} No & \num{-0.313}  & (\num{0.061}) & {}[\num{0.000}]& \{0.000\} \\
\\
\textbf{Mother Education Level}\\
\hphantom{a} Less than High School & \num{-0.106}  & (\num{0.047})  & {}[\num{0.026}]& - \\
\hphantom{a} High School & \num{-0.190}  & (\num{0.059}) & {}[\num{0.002}]& \{0.215\} \\
\hphantom{a} Some College and more & \num{-0.313}  & (\num{0.063})  & {}[\num{0.000}]& \{0.001\} 
\\ 
\\
\textbf{Family Income}\\
\hphantom{a} Level 1 & \num{-0.117}  & (\num{0.052})  & {}[\num{0.028}]& - \\
\hphantom{a} Level 2 & \num{-0.186}  & (\num{0.062}) & {}[\num{0.003}]& \{0.256\} \\
\hphantom{a} Level 3 & \num{-0.306}  & (\num{0.054}) & {}[\num{0.000}]& \{0.003\} \\
 \addlinespace
\hline \hline 
      \end{tabular}
\begin{tablenotes}
\item \footnotesize{Notes: The table presents the estimates of $\beta$ from equation \ref{res:main_results} for students in UFBA interacted with several different variables. I multiply the coefficient by minus one, therefore we can see the coefficients as a decrease of 100\% of the weighted hiring measure. The recession between 2014-2016 reduced the weighted hiring measure by 30\%. Each group of heterogeneity is estimated separately. The first column shows the point estimates, the second column the standard error and the third column the p-value of the test whether the effect is different than zero. The fourth column presents the tests that each level of the heteogeneity analyzed is the same as the first one presente in each group. Standard errors are clusteres at the major level. All regressions include fixed effects for program, time of admission, fall semester, and demographic cells (gender, race and age).}
\end{tablenotes}
\end{threeparttable}
\end{adjustbox}
\end{table}

\section{Conclusion}

College students graduating in a recession face negative and persistent effects on their labor market careers. In this paper, I investigate whether college students delay graduation to avoid entering a depressed labor market. I explore variations in the labor market conditions over time, geography, and majors in Brazil. I find that students in public institutions are 2.1pp less likely to graduate on time when facing a recession that decreases new hires by 30\%. This represents an increase in the average duration by 0.11 semesters, or 1 out of 18 students delaying by one year. I do not find any effect for students in private institutes. 

The delaying effects are more prominent for students with higher scores, in higher-earning majors, and from more advantaged backgrounds. This has important implications for inequality since students with better resources can better shield themselves from labor market fluctuations by controlling the time of graduation. The results show how the institutional setting is important when assessing college students' response to labor market conditions. 

\singlespacing
\nocite{cutler2015,borgschulte2018,fernandez2018}
\bibliography{main.bib}


\clearpage
\appendix
\section*{Online Appendices}

\doublespacing 
\addcontentsline{toc}{section}{Appendices}
\titleformat{\subsection}{\normalfont\fontsize{12}{12} \bfseries \center }{Appendix \Alph{subsection} - }{0em}{}
\renewcommand{\thesubsection}{\Alph{subsection}}
\titlespacing{\subsubsection}{0pt}{0cm}{0cm}

\setcounter{table}{0}
\renewcommand\thetable{A.\arabic{table}}

\setcounter{figure}{0}
\renewcommand\thefigure{A.\arabic{figure}}

\subsection{Additional Exercises}

\subsubsection{Reweighting}\label{app:reweigthing}

Public and private institutions differ in several characteristics, such as course quality, selectivity, and student composition. In this exercise, I reweight the observations from the private institutions to have the same distribution as the students from the public universities. The first column of table \ref{tab:reweigthing} shows the baseline result. In the second column, I reweight the observations from the private institutions to have the same distribution of majors as the public universities. The reweight procedure yields the same proportion of students in each major for every semester in the sample. The numbers of observations are slightly different because if no private universities are offering a major in a given semester, the students from public institutions from that major-year are dropped. 

\begin{table}[!ht] \centering 
    \begin{adjustbox}{max width = \textwidth, width = \textwidth, center}
    \begin{threeparttable}
    \caption{Reweighting}\label{tab:reweigthing}
    \begin{tabular}{lcccccc} 
    \\[-1.8ex]\hline 
    \hline \\[-1.8ex]
    \multicolumn{1}{r}{\emph{Outcome:}} & \multicolumn{6}{c}{\emph{On-time graduation}}\\
    \addlinespace
    & (1) & (2) & (3) & (4) & (5) &(6)\\\addlinespace
    Reweighting for: & Benchmark & Majors & Demographics & \% Students& Quality& Quality \\
    &&&&Public HS & (ENADE) & (CPC)\\[.8ex] \hline \\[-.8ex]
    \addlinespace
    Hiring x Public & \num{-0.070} & \num{-0.067} & \num{-0.068} & \num{-0.064} & \num{-0.059} & \num{-0.059}\\
     (s.e.) & (\num{0.031}) & (\num{0.033}) & (\num{0.033}) & (\num{0.036}) & (\num{0.035}) & (\num{0.035})\\
     
     [p-value] & {}[\num{0.032}] & {}[\num{0.053}] & {}[\num{0.050}] & {}[\num{0.082}] & {}[\num{0.107}] & {}[\num{0.108}]\\
     \\
    Hiring x Private & \num{-0.003} & \num{0.000} & \num{-0.005} & \num{0.000} & \num{-0.016} & \num{-0.016}\\
     (s.e.) & (\num{0.040}) & (\num{0.032}) & (\num{0.035}) & (\num{0.039}) & (\num{0.038}) & (\num{0.038})\\
     
     [p-value] & {}[\num{0.934}] & {}[\num{0.991}] & {}[\num{0.886}] & {}[\num{0.995}] & {}[\num{0.676}] & {}[\num{0.683}]\\
    \midrule
    N Obs & {4,058,758} & {4,016,188} & {4,058,743} & {3,452,328} & {3,600,588} & {3,596,268}\\
    \\
    p-value ($\beta_{\text{public}}=\beta_{\text{private}}$) & \{0.084\}& \{0.058\}& \{0.099\}& \{0.159\}& \{0.350\}& \{0.352\}\\
    \\
    ratio $({\beta_{\text{private}}}/{\beta_{\text{public}}})$ & 0.048&0.006&0.075&0.004&0.273&0.267\\
    \hline 
    \hline \\[-1.8ex] 
    \end{tabular} 
    \begin{tablenotes}
    \item \footnotesize{Notes: The table presents the estimation of $\beta$ from equation \ref{res:main_results} interacted with an indicator for students belonging to a public or private institution. I multiply the coefficient by minus one, therefore, we can see the coefficients as a decrease of 100\% of the weighted hiring measure. The recession between 2014-2016 reduced the weighted hiring measure by 30\%. The first column presents the baseline result from Table \ref{res:main_results}. Each of the following columns reweights the observations from the private institutions to have the same distribution as the public universities according to a different dimension. The reweighting procedure is described in the Appendix \ref{app:reweigthing}. Each exercise is computed separately. Standard errors are clustered at both the major and state levels (\emph{two-way clustering}). All regressions include fixed effects for program, major-admission time, fall semester, and demographic cells (gender, race, and age). The last but one row shows the p-value testing whether the effect is the same for public and private institutions. The last row shows the ratio of the point estimates for public and private effects.}
    \end{tablenotes}
    \end{threeparttable}
    \end{adjustbox}
\end{table} 

The third column reweights for the students' demographics jointly defined by gender, race, and age of entry. The fourth column matches the sample by comparing the proportion of students that studied in public high schools in a given program. For this exercise, I compute the proportion of students in public schools at every 5\% (0-5\%, 5-10\%, 10-15\%, $\hdots$,95\%-100\%). The last two columns reweight the distributions according to two measures of quality: the average score at the ENADE exam (5th column) and the CPC measure (6th column). In both cases, I computed qualities rounding at the one-digit level. 

\subsubsection{Tobit Regression}\label{app:tobit}

I run a Tobit regression where the outcome is the censored semester of actual graduation. Therefore Student $i$ that graduated on time will have $Y_i=0$, while Student $j$ that graduated 3 semesters after their expected graduation date will have $Y_j=3$. A Student $k$ that did not graduate until the maximum observed semester (for instance, 8) for their cohort will have the \emph{censored} $Y_k=8$. 

I cannot run the equivalent of equation \ref{eq:main_specification} because I do not have enough computational power to include the thousands of fixed effects present in my baseline specification. Instead, I run the following specification: 
\begingroup
\large
\[ Y_i = MWH_{t_1(i),m(i),s(i)}+\eta_{t_0(1)}+\nu_{s(i)}+\gamma X_i+\delta Z_{p(i)}+\varepsilon_i\]
\endgroup

In this specification, I substitute the program fixed effects with both state fixed effects and program characteristics (major, schedule, duration, and others). The results of the Tobit regression are presented in the table \ref{res:app_tobit} below. Another limitation is the estimation of standard errors, which can be done only for one cluster variable. I chose to estimate with the state level because it is the variable with the lowest number of groups (compared to the number of majors). 

\begin{table}[!htbp] \centering 
\begin{adjustbox}{max width = \textwidth, width = .5\textwidth, center}
\begin{threeparttable}
\caption{Censored regression}\label{res:app_tobit}
\begin{tabular}{lcc} 
\\[-1.8ex]\hline 
\hline \\[-1.8ex]
\multicolumn{1}{r}{\emph{Outcome:}} & \multicolumn{2}{c}{\emph{Semester of graduation}}\\
\addlinespace
& (1) & (2) \\[.8ex] \hline \\[-.8ex]
Hiring & \num{0.455} & \num{0.353}\\
 (s.e.) & (\num{0.504}) & (\num{0.508})\\
 
 [p-value] & [\num{0.367}] & [\num{0.488}]\\
\\
Num.Obs. & {1,260,875} & {1,024,315}\\
\\
\hline\addlinespace
Admission year FE & $\checkmark$& $\checkmark$\\
State FE & $\checkmark$& $\checkmark$\\
Demographics & $\checkmark$& $\checkmark$\\
Major FE & $\checkmark$& $\checkmark$\\
Schedule and Duration & $\checkmark$ & $\checkmark$\\
Program Characteristics & - & $\checkmark$\\
\hline 
\hline \\[-1.8ex] 
\end{tabular} 
\begin{tablenotes}
\item \footnotesize{Notes: Tobit regression using as the outcome the censored semester of graduation relative to expected graduation for students in public universities. I multiply the coefficient by minus one, therefore, we can see the coefficients as a decrease of 100\% of the weighted hiring measure. The recession between 2014-2016 reduced the weighted hiring measure by 30\%. Each group of heterogeneity is estimated separately. Standard errors are clustered at the state level. Each column differs only by the set of included controls.}
\end{tablenotes}
\end{threeparttable}
\end{adjustbox}
\end{table} 

\subsubsection{Sensitivity Analysis}\label{app:sensitivity}

In order to reliable compute the major-weighted hiring measure, I apply some sampling restrictions. I detail each of them below:

\begin{enumerate}
 \item \underline{\textbf{RAIS tolerance (20\%)}} --- First, I drop occupation codes whose maximum error rate across all years between 2003-2019 are above 20\%. This is likely to indicate issues with the structure of the coding scheme of occupation to have changed in my sampling period. 
 \item \underline{\textbf{\# observations in each Census (20)}} --- I only keep majors that I observe at least 20 individuals working in each of the Demographic Censuses (2000 and 2010). 
 \item \underline{\textbf{\% Observed (30\%)}} --- I drop majors that I only observed fewer than 30\% of the individuals working in occupations that were not discarded in procedure 1 above. 
 \item \underline{\textbf{Maximum Age (40)}} --- I only use individuals aged between 20 and 40 to compute the occupation weights in order to capture occupations that are relevant for recent college graduates. 
\end{enumerate}

Table \ref{res:sensitivity} below assess how the baseline result changes when we vary each of the above-specified thresholds. We can see that the main results are not sensitive to any sampling decision. The table also presents the effective number of majors and occupations and the percentual of employment covered for each sampling decision. 

\begin{table}[!ht] \centering 
    \begin{adjustbox}{max width = \textwidth, width = \textwidth, center}
    \begin{threeparttable}
    \caption{Sensitivity Analysis}\label{res:sensitivity}
    \begin{tabular}{lccccccccc} 
    \\[-1.8ex]\hline 
    \hline \\[-1.8ex]
    \multicolumn{10}{l}{\emph{Outcome: On-time graduation}}\\
    \addlinespace
    &Benchmark & \multicolumn{2}{c}{RAIS tolerance} & \multicolumn{2}{c}{\# obs Census} & 
    \multicolumn{2}{c}{\% observed} & \multicolumn{2}{c}{Max Age} \\ 
    \cmidrule(l{5pt}r{5pt}){2-2}
    \cmidrule(l{5pt}r{5pt}){3-4}
    \cmidrule(l{5pt}r{5pt}){5-6}
    \cmidrule(l{5pt}r{5pt}){7-8}
    \cmidrule(l{5pt}r{5pt}){9-10}
    \addlinespace
    & - & 0.15 & 0.25 & 0 & 40 & 0.00 & 0.50 & 35 & 45  \\[.8ex] \hline \\[-.8ex]
    \addlinespace
    Hiring x Public & \num{-0.070} & \num{-0.051} & \num{-0.073} & \num{-0.069} & \num{-0.070} & \num{-0.091} & \num{-0.061} & \num{-0.072} & \num{-0.069}\\
 (s.e.) & (\num{0.031}) & (\num{0.026}) & (\num{0.032}) & (\num{0.030}) & (\num{0.031}) & (\num{0.036}) & (\num{0.028}) & (\num{0.031}) & (\num{0.030})\\
 
 [p-value]& {}[\num{0.032}] & {}[\num{0.059}] & {}[\num{0.029}] & {}[\num{0.031}] & {}[\num{0.032}] & {}[\num{0.019}] & {}[\num{0.035}] & {}[\num{0.031}] & {}[\num{0.032}]\\
 \\
Hiring x Private & \num{-0.003} & \num{-0.011} & \num{-0.004} & \num{-0.003} & \num{-0.004} & \num{-0.018} & \num{-0.005} & \num{-0.003} & \num{-0.004}\\
 (s.e.)& (\num{0.040}) & (\num{0.040}) & (\num{0.039}) & (\num{0.040}) & (\num{0.040}) & (\num{0.044}) & (\num{0.040}) & (\num{0.039}) & (\num{0.040})\\
 
 [p-value] & {}[\num{0.934}] & {}[\num{0.775}] & {}[\num{0.917}] & {}[\num{0.935}] & {}[\num{0.920}] & {}[\num{0.683}] & {}[\num{0.899}] & {}[\num{0.944}] & {}[\num{0.926}]\\
\midrule
N Obs & {4,058,758} & {2,811,362} & {4,124,702} & {4,071,642} & {4,051,776} & {4,653,332} & {3,895,403} & {4,056,786} & {4,058,758}\\
\\
    \# of Majors & 64 & 55 & 65 & 79 & 62 & 69 & 62 & 63 & 65 \\
    \# of Occupations & 143 & 106 & 150 & 143 & 143 & 143 & 143 & 143 & 143  \\
    \% employment & 0.925 & 0.725 & 0.941 & 0.925 & 0.925 & 0.925 & 0.925 & 0.925 & 0.925 \\
    \hline 
    \hline \\[-1.8ex] 
    \end{tabular} 
    \begin{tablenotes}
    \item \footnotesize{Notes: The table presents the estimation of $\beta$ from equation \ref{res:main_results} interacted with an indicator for students belonging to a public or private institution. I multiply the coefficient by minus one, therefore we can see the coefficients as a decrease of 100\% of the weighted hiring measure. The recession between 2014-2016 reduced the weighted hiring measure by 30\%. The first column presents the baseline result from Table \ref{res:main_results}. Each of the next columns reestimate our main regression with a sample that was computed varying of my sampling decision rules. Each exercise is computed separately. Standard errors are clusteres at both the major and state levels (\emph{two-way clustering}). All regressions include fixed effects for program, time of admission, fall semester, and demographic cells (gender, race and age).}
    \end{tablenotes}
    \end{threeparttable}
    \end{adjustbox}
\end{table} 

I also compute the main results excluding the years where the individual information was linked over time using the matching algorithm instead of the unique identifiers. Table \ref{res:rob_2017} below presents these results, that are really similar to the main results presented in the table \ref{res:main_results}.

\begin{table}[!ht] \centering 
    \begin{adjustbox}{max width = \textwidth, width = \textwidth, center}
    \begin{threeparttable}
    \caption{Results restricting the data to years that unique identifiers are observed}\label{res:rob_2017}
    \begin{tabular}{lcccccc} 
    \\[-1.8ex]\hline 
    \hline \\[-1.8ex]
    \multicolumn{1}{r}{\emph{Outcome:}} & \multicolumn{6}{c}{\emph{On-time graduation}}\\
    \addlinespace
    & (1) & (2) & (3) & (4) & (5) & (6) \\[.8ex] \hline \\[-.8ex]
    \addlinespace
     Hiring & \num{-0.012} &  &  &  &  & \\
     & (\num{0.021}) &  &  &  &  & \\
     & {}[\num{0.585}] &  &  &  &  & \\
    Hiring x Public &  & \num{-0.060} & \num{-0.053} & \num{-0.053} & \num{-0.059} & \num{-0.048}\\
     &  & (\num{0.028}) & (\num{0.028}) & (\num{0.028}) & (\num{0.032}) & (\num{0.027})\\
     &  & {}[\num{0.043}] & {}[\num{0.066}] & {}[\num{0.070}] & {}[\num{0.072}] & {}[\num{0.092}]\\
    Hiring x Private &  & \num{0.004} & \num{0.010} & \num{0.013} & \num{-0.006} & \num{0.007}\\
     &  & (\num{0.030}) & (\num{0.025}) & (\num{0.024}) & (\num{0.032}) & (\num{0.029})\\
     &  & {}[\num{0.905}] & {}[\num{0.684}] & {}[\num{0.586}] & {}[\num{0.858}] & {}[\num{0.800}]\\
    \midrule
    N Obs & {3,121,188} & {3,121,188} & {3,121,188} & {3,121,188} & {3,121,188} & {3,121,188}\\    
    \\
    p-value ($\beta_{\text{public}}=\beta_{\text{private}}$) & - & \{0.100\} & \{0.022\} & \{0.028\} & \{0.113\} & \{0.101\} \\ 
    \\
    Major-State FE &$\checkmark$ & $\checkmark$ &- & - & - & - \\
    Program FE & -&-& $\checkmark$& $\checkmark$& $\checkmark$ & $\checkmark$\\
    Demographics & -& - & $\checkmark$& $\checkmark$&$\checkmark$ & $\checkmark$\\
    \addlinespace
    \multirow{2}{*}{Time Trend} & \multirow{2}{*}{Quadratic} & \multirow{2}{*}{Quadratic} & \multirow{2}{*}{Quadratic} & \multirow{2}{*}{Quadratic} & Admission & Major-Admission \\
    &&&&&Time FE & Time FE\\
    \hline 
    \hline \\[-1.8ex] 
    \end{tabular} 
    \begin{tablenotes}
    \item \footnotesize{Notes: The table presents the estimation of $\beta$ from equation \ref{res:main_results} for the data restricted to the year 2017 that does not rely on the matching algorithm to follow students. I multiply the coefficient by minus one, therefore, we can see the coefficients as a decrease of 100\% of the weighted hiring measure. The recession between 2014-2016 reduced the weighted hiring measure by 30\%. The first column presents the overall results. In columns 2-6, I interact all variables with an indicator of whether the students belong to a private or public institution. Columns 2-6 differ in the set of control variables included in each specification. Standard errors are clustered at both the major and state levels (\emph{two-way clustering}). The p-value of the test whether the effect for public and private institutions are the same is provided for each specification. FE stands for fixed effects.}
    \end{tablenotes}
    \end{threeparttable}
    \end{adjustbox}
\end{table}

\subsubsection{Sample of all students first enrolled in the program}\label{sample_restriction}

Table \ref{res:robustness_freshmen} below reproduces the main estimates from Table~\ref{res:main_results} using the sample of all students first enrolled in the program, irrespectively of their enrollment status in the year of expected graduation.

\begin{table}[!ht] \centering 
    \begin{adjustbox}{max width = \textwidth, width = \textwidth, center}
    \begin{threeparttable}
    \caption{Robustness: sample all students enrolled}\label{res:robustness_freshmen}
    \begin{tabular}{lcccccc} 
    \\[-1.8ex]\hline 
    \hline \\[-1.8ex]
    \multicolumn{1}{r}{\emph{Outcome:}} & \multicolumn{6}{c}{\emph{On-time graduation}}\\
    \addlinespace
    & (1) & (2) & (3) & (4) & (5) & (6) \\[.8ex] \hline \\[-.8ex]
    \addlinespace
    Hiring & \num{0.004} &  &  &  &  & \\
    & (\num{0.019}) &  &  &  &  & \\
    & {}[\num{0.846}] &  &  &  &  & \\
    \\
   Hiring x Public &  & \num{-0.061} & \num{-0.043} & \num{-0.049} & \num{-0.036} & \num{-0.036}\\
    &  & (\num{0.019}) & (\num{0.016}) & (\num{0.016}) & (\num{0.015}) & (\num{0.017})\\
    &  & {}[\num{0.003}] & {}[\num{0.011}] & {}[\num{0.006}] & {}[\num{0.023}] & {}[\num{0.050}]\\
    \\
   Hiring x Private &  & \num{0.040} & \num{0.041} & \num{0.041} & \num{0.012} & \num{0.030}\\
    &  & (\num{0.027}) & (\num{0.026}) & (\num{0.026}) & (\num{0.023}) & (\num{0.026})\\
    &  & {}[\num{0.152}] & {}[\num{0.129}] & {}[\num{0.131}] & {}[\num{0.625}] & {}[\num{0.273}]\\
 \\
    \midrule
    N Obs & \num{6595407} & \num{6595407} & \num{6595407} & \num{6595407} & \num{6595407} & \num{6595407}\\
    \\
    p-value ($\beta_{\text{public}}=\beta_{\text{private}}$) & - & \{0.001\} & \{0.003\} & \{0.003\} & \{0.050\} & \{0.024\} \\ 
    \\
    Major-State FE &$\checkmark$ & $\checkmark$ &- & - & - & - \\
    Program FE & -&-& $\checkmark$& $\checkmark$& $\checkmark$ & $\checkmark$\\
    Demographics & -& - & $\checkmark$& $\checkmark$&$\checkmark$ & $\checkmark$\\
    \addlinespace
    \multirow{2}{*}{Time Trend} & \multirow{2}{*}{Quadratic} & \multirow{2}{*}{Quadratic} & \multirow{2}{*}{Quadratic} & \multirow{2}{*}{Quadratic} & Admission & Major-Admission \\
    &&&&&Time FE & Time FE\\
    \hline 
    \hline \\[-1.8ex] 
    \end{tabular} 
    \begin{tablenotes}
    \item \footnotesize{Notes: The table presents the estimation of $\beta$ from equation \ref{res:main_results} considering the sample of all students enrolled in the first semester of the program, irrespective of whether they were still enrolled in the year of expected graduation. I multiply the coefficient by minus one, therefore, we can see the coefficients as a decrease of 100\% of the weighted hiring measure. The recession between 2014-2016 reduced the weighted hiring measure by 30\%. The first column presents the overall results. In columns 2-6, I interact all variables with an indicator of whether the students belong to a private or public institution. Columns 2-6 differ in the set of control variables included in each specification. Standard errors are clustered at both the major and state levels (\emph{two-way clustering}). The p-value of the test whether the effect for public and private institutions are the same is provided for each specification. FE stands for fixed effects.}
    \end{tablenotes}
    \end{threeparttable}
    \end{adjustbox}
\end{table}

In this sample, only 58.2\% continue to be enrolled in their year of expected graduation for public institutions and 45.6\% for private institutions. The on-time graduation rate for this sample is 19.1\% for students in public universities and 23.4\% for students in private institutes. These numbers implies that a recession of the size of the 2014--2016 recession reduces the on-time graduation in public universities by $(0.30*0.036)/0.191=5.6\%$, compared to 6.5\% from the benchmark estimates.

\clearpage
\subsection{Data}\label{data_details}

\subsubsection{Higher Education Census}

I use all the Higher Education Census (\emph{Censo da Educação Superior}) from 2009 to 2019. From 2009 to 2017, the data includes unique identifiers at the student and enrollment level, allowing me to follow individual students over time. For 2018 and 2019, these identifiers are suppressed from the public files. 

I develop an algorithm to match students across time using the information on the university, major, first enrollment year, date of birth, gender, and place of birth when this information yields a unique observation. To test the algorithm, I apply it to the years before 2017, and I obtain a success rate between 91.7\% and 94.3\%. Notably, out of the 5.7\%-8.3\% not successfully matched, less than 0.01\% are from two distinct students being incorrectly matched. The majority of them are students that I do not attempt to match because they do not have unique observations across the variables used in the algorithm. I always match only two consecutive years to maximize the algorithm's success rate. 

After obtaining the unique identifiers for the entire period, I define enrollment as a combination of student-university-major. The enrollment date and schedule are obtained from the first time they appear in the data.\footnote{Some years have missing data for the period of study. In these years, I consider the first non-missing information.}. I obtain personal information for each student, including date of birth, gender, race, and place of birth as the mode for each variable across all years. 

I then apply several cleaning procedures to result in a homogeneous sample. I start with a universe of 40 million unique enrollments from 30 million students. I then drop observations where I could not obtain key characteristics from their programs. I lose 29\% of students in this step, mainly those who first enrolled much earlier than 2009. I drop students that are not enrolled in B.A. equivalent programs (7.9\%), not enrolled in in-person programs (0.05\%), with no well-defined duration (0.6\%), and with first enrollment before 2004 (0.6\%).\footnote{The minimum duration of a bachelor's course is three years, according to Ministry of Education Resolution 2 from 2007.} Lastly, I drop students that were not between 17 and 22 years old when they first enrolled, dropping 23.2\% of the sample and with expected graduation after 2019 (dropping 11.5\%). The final sample has 7.8 million unique students, 9.5 million enrollments in 2,342 institutes, 74 majors, and 40,849 programs (major-institution-schedule). 

\subsubsection{UFBA}

I obtain panel data from all students enrolled in the university between 2003 and 2017 from the Federal University of Bahia (UFBA). In the cleaning procedure, I discard duplicated observations at the student-semester-course-status level, and I remove the fall of 2004 coursework since all classes were repeated in the spring due to a strike.

The initial data set covers 80,165 unique students. I remove students with conflicting data for the admission date (0.001\%), that I cannot obtain information on the program (1.6\%), with expected graduation after 2019 (20.2\%), and in programs with less than 3-year duration (0.5\%). The resulting sample has 62,245 unique students enrolled in 230 programs (major-schedule). 

For every semester, I check whether students were working in the matched employer-employee data set, classifying ``working" as having worked for at least one month during the semester. 

\subsubsection{Household Surveys (PNAD and PNADC)}

I use the Brazilian National Household Survey (PNAD, \emph{Pesquisa Nacional por Amostra de Domicílios}) from 2002-2009 and 2011, and the Continuous Brazilian National Household Survey (PNADC, \emph{Pesquisa Nacional por Amostra de Domicílios Contínua}) from 2012-2019, to compute the state unemployment rate. I restrict the sample to individuals aged 27 and 65 and estimate the unemployment rate using the sampling structure of the surveys (strata and sampling weights). 

\subsubsection{Demographic Population Census}

I use data from the two demographic censuses in 2000 and 2010 (\emph{Censo Demográfico})throughout my analysis. First, I use the 2010 Census to obtain the state unemployment rates, with the same sampling restrictions as in the household surveys.\footnote{In the year that the demographic census is collected, the household surveys are not collected.}

I also use the two censuses to obtain the occupation-major weights. To ensure that I am calculating these weights according to the market new graduates face, I restrict the sample to working, college-educated 20-40-year-olds. I drop majors with less than 20 individuals working in any of the two Censuses or where more than 70\% of individuals work in occupations not covered in the study. These restrictions remove 20 majors from the 84 majors listed. I obtain the weights taking into account the Census sampling weights. 

Lastly, I use the two censuses to obtain the average earnings of each major, considering individuals between 27 and 65 years old, working, and with a college degree in a given major. I also obtain the non-major-specific average earnings for each municipality using all working individuals aged 27-65 years. I use the total labor earnings deflated by the consumer price index (IPCA) for both measures. I remove composition effects by residualizing earnings from gender, race, and age. 

\subsubsection{Matched Employer-employee (RAIS)}

For most of the analysis, I use the public version of the matched employer-employee data, available in the data-leak \emph{Base dos Dados}.\footnote{https://basedosdados.org} from 2003-2019. I restrict the sample to individuals between 27 and 65 years of age. There were initially 176 occupations at the 4-digit code level. I remove occupations that are likely to be affected by the modifications in the occupation codes between 2008-2012. In order to assess that, I compute the stock of employees working in a given occupation on December 31st of year $t$ ($\text{stock}_t$). I then compute the flow of employees in this occupation (hires - layoffs) in January of $t+1$ ($\text{flow}_{t+1}$) and the stock of employees at the end of this month --- ($\text{stock}_{t+1}$). In the absence of codification changes and measurement errors, we expect:
\[\text{stock}_{t+1} = \text{stock}_t+\text{flow}_{t+1}\]
I compute the maximum proportion error for each occupation $o$ as 
\[ \text{Max}_o = \max_t\left\{\left|\frac{\text{stock}^o_{t+1} - \text{flow}^o_{t+1}}{\text{stock}^o_{t}}\right|\right\}\]
I then drop occupations whose maximum error is above 20\% for any given year. 

For the UFBA data, I use the restricted identified version of the data, available for 2003-2018, and I apply a similar procedure.

\subsubsection{ENADE}

I obtain the ENADE (\emph{Exame Nacional de Desempenho dos Estudantes}) data from 2004 to 2019. I restrict the sample to individuals who answered at least one question from those I selected (working status, parents' educational level, type of high school). I remove programs in which, across all years, there were fewer than 30 valid responses or fewer than 50\% of students had valid answers. For the ENADE and CPC scores, I use the final scores released by the INEP agency.

\clearpage
\subsection{Appendix Table and Figures}

\begin{figure}[!ht]
    \centering
    \caption{Distribution of scores by public and private schools}\label{fig:education_setting}
    \includegraphics[width=\textwidth]{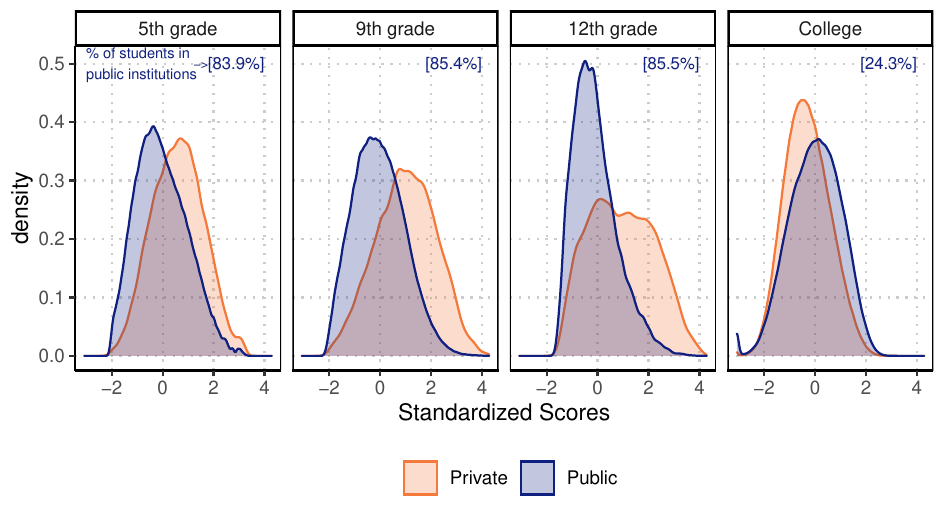}
    \caption*{\footnotesize{Notes: The figure shows the density of standardized scores for public (in \textcolor{blue_graph}{blue}) and private schools (in \textcolor{orange_graph}{orange}). The scores for students in 5th, 9th, and 12th (3rd grade of High School) grades were obtained from the national SAEB exam in 2015. The grades for college students use the ENADE exam between 2014--2016. All scores were normalized to have zero mean and one standard deviation for the public schools. The numbers in brackets show the proportion of students in public schools/universities.}}
\end{figure}

\begin{figure}[!ht]
    \centering
        \caption{Proportion of students by program duration}\label{fig:major_length}
    \includegraphics[width=\textwidth]{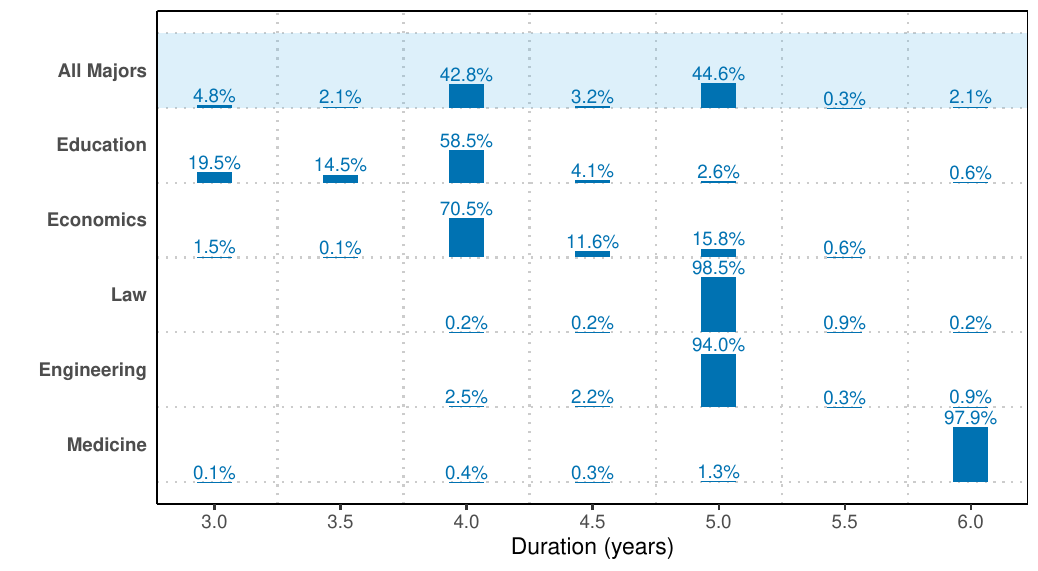}
    \caption*{\footnotesize{Notes: In each row, the figure shows the proportion of students enrolled in programs with different duration, from 3 to 6 years. In the first row is the distribution for all students enrolled in higher education. In the next five rows, I present the distribution for students majoring in Education, Economics, Law, Engineering, and Medicine. The numbers were calculated using the Higher Education Census from 2009 to 2019.}}
\end{figure}

\begin{figure}[!ht]
    \centering
    \caption{Occupation-major weights - Reliability and uniqueness}\label{fig:weights_assessment}
        \includegraphics[width=\textwidth]{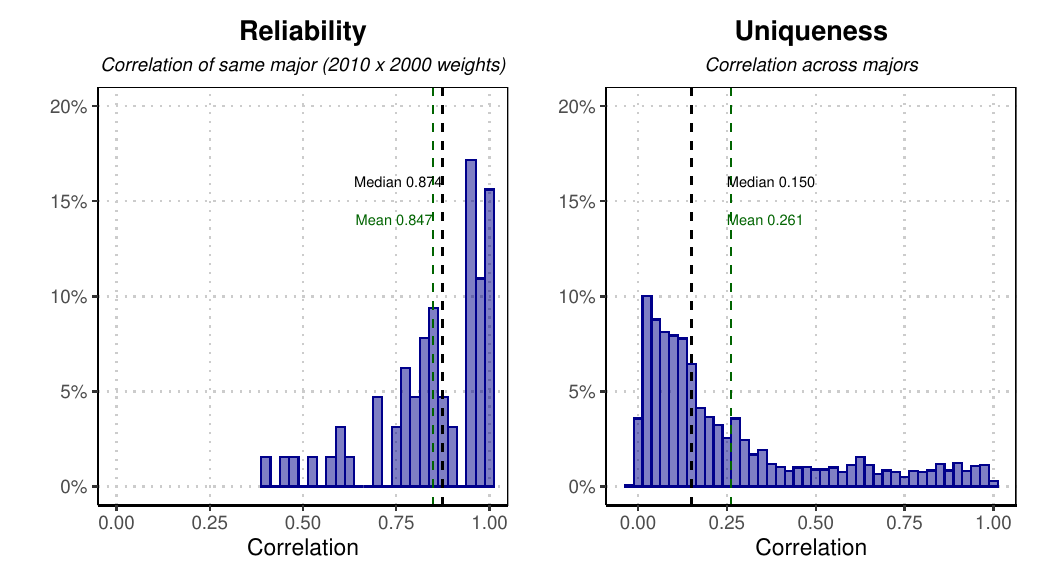}
    \caption*{\footnotesize{Notes: The panel on the left shows the histogram for the correlation of the vector of weights in 2000 and 2010 for the same major in the main sample. The vertical \textbf{black} line shows the median value and the \textcolor{green}{green} line the mean. The panel on the right shows the histogram for the comparison of the vector of weights across different majors.}}
\end{figure}

\begin{figure}[!ht]
    \centering
    \caption{Occupation-major weights - reliability and Uniqueness (UFBA sample)}\label{fig:weights_assessment_UFBA}
        \includegraphics[width=\textwidth]{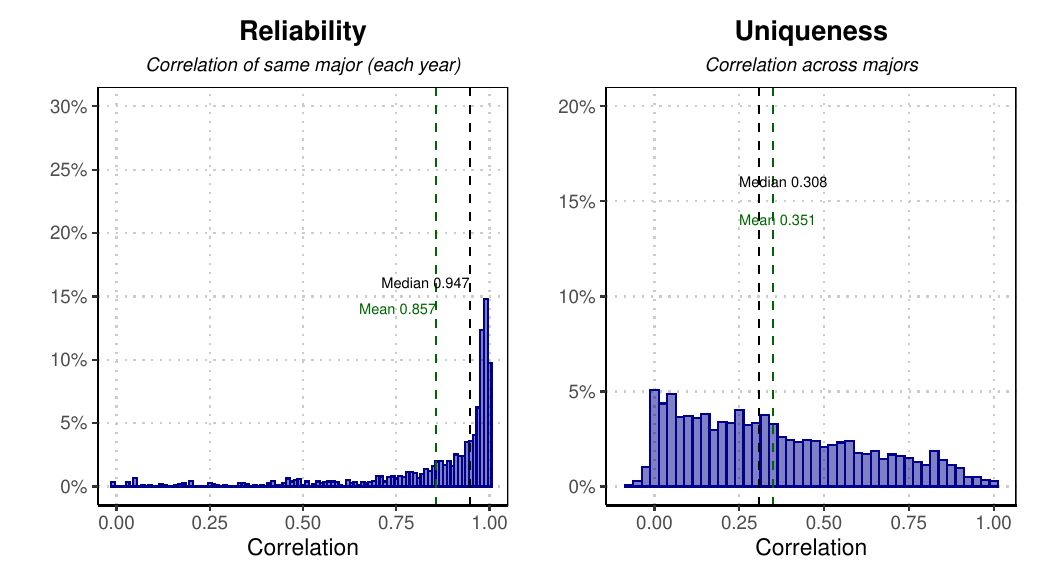}
    \caption*{\footnotesize{Notes: The panel on the left shows the histogram for the correlation of the vector of weights in different years for the same major in the UFBA sample. The vertical \textbf{black} line shows the median value and the \textcolor{green}{green} line the mean. The panel on the right shows the histogram for the comparison of the vector of weights across different majors.}}
\end{figure}

\begin{figure}[!ht]
    \centering
    \caption{Heterogeneity by Major}\label{fig:het_major}
    \includegraphics[width=.825\textwidth]{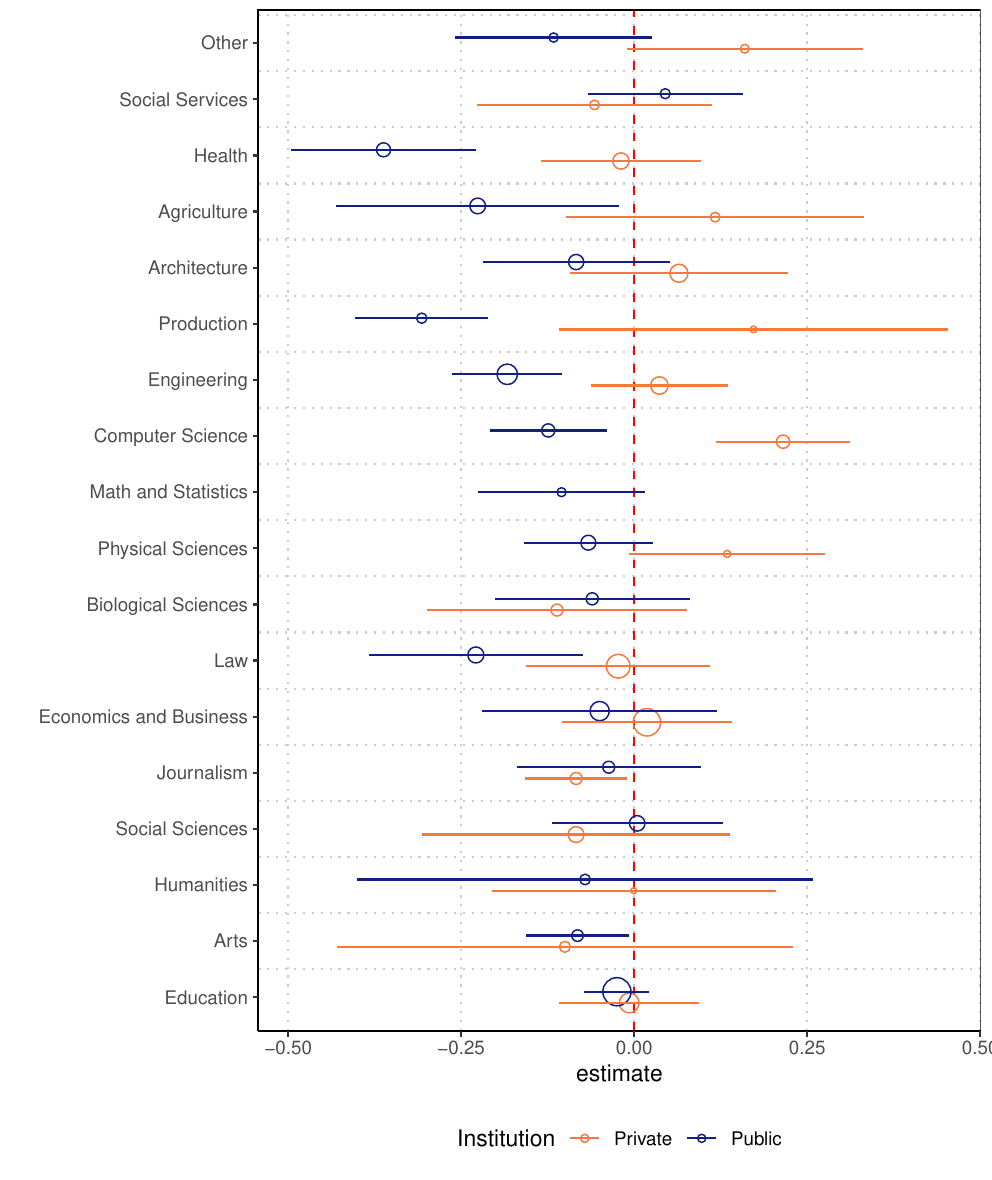}
    \caption*{\footnotesize{Notes: The figure presents the estimation of $\beta$ from equation \ref{res:main_results} interacted with an indicator for students belonging to a public or private institution and major groups. I multiply the coefficient by minus one, therefore, we can see the coefficients as a decrease of 100\% of the weighted hiring measure. The recession between 2014--2016 reduced the weighted hiring measure by 30\%. The circles represent the point estimates, and the lines the 95\% confidence intervals. Standard errors are clustered at the major and state levels (\emph{two-way clustering}). All regressions include fixed effects for program, time of admission, fall semester, and demographic cells (gender, race, and age). The \textcolor{orange}{orange} color represents estimates for private institutions and \textcolor{blue}{blue} color for public universities. All majors with at least 1,000 students are displayed in the figure.}}
\end{figure}

\begin{figure}[!ht]
    \centering
    \caption{Placebo Effects}\label{fig:placebo}
    \includegraphics[width=.9\textwidth]{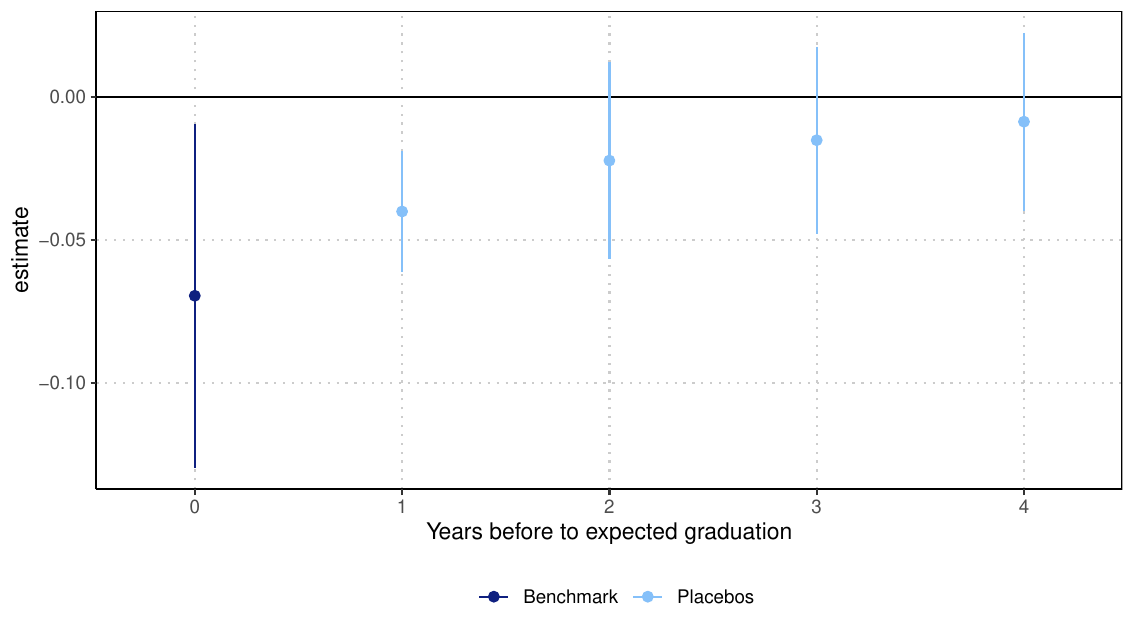}
    \caption*{\footnotesize{Notes: The figure presents the estimation of $\beta$ from equation \ref{res:main_results} for students in public universities. I multiply the coefficient by minus one, therefore, we can see the coefficients as a decrease of 100\% of the weighted hiring measure. The recession between 2014--2016 reduced the weighted hiring measure by 30\%. The circles represent the point estimates, and the lines the 95\% confidence intervals. Each regression uses the major-weighted hiring measure for $\tau$ years before the expected graduation, for $\tau\in[0,4]$. Standard errors are clustered at the major and state levels (\emph{two-way clustering}). All regressions include fixed effects for program, time of admission, fall semester, and demographic cells (gender, race, and age).}}
\end{figure}

\begin{figure}[!ht]
    \centering
    \caption{Residualized On-time Graduation and MWH}\label{fig:np}
        \includegraphics[width=\textwidth]{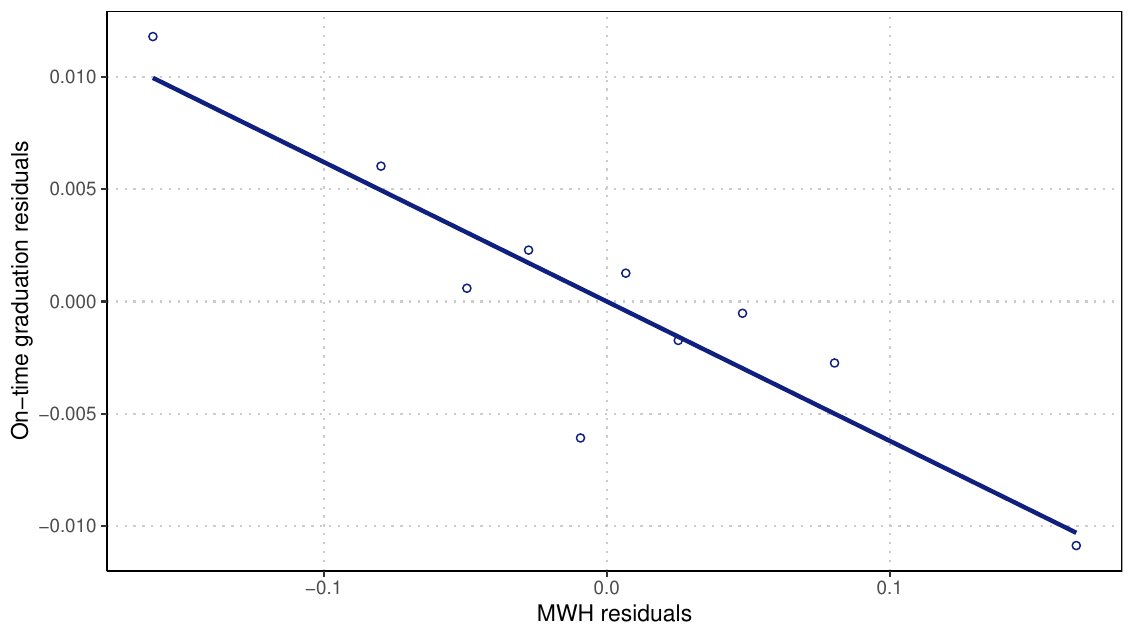}
    \caption*{\footnotesize{Notes: The circles represent the binned averages of the residualized on-time graduation dummy and the residualized major-weighted hiring measure, dividing the sample into 10 equal-sized groups ordered by the hiring measure residuals. The residuals are obtained regressing each variable on the full set of fixed effects (semester, program, time of admission, and demographic cells --- gender, race, and age). The line is the slope ($\beta$ coefficient from equation \ref{eq:main_specification}) of the benchmark OLS specification.}}
\end{figure}

\begin{table}[!ht] \centering 
    \begin{adjustbox}{max width = \textwidth, width = .6\textwidth, center}
    \begin{threeparttable}
    \caption{Balance}\label{tab:balance}
    \begin{tabular}{lccc} 
    \\[-1.8ex]\hline 
    \hline \\[-1.8ex]
    \multicolumn{1}{r}{\emph{Outcome:}} & Women & Black/Native & Age at Entry\\
    \addlinespace
    & (1) & (2) & (3)\\[.8ex] \hline \\[-.8ex]
    \addlinespace
    Hiring & \num{-0.001} & \num{0.003} & \num{0.075}\\
 
 (s.e.)   & (\num{0.007}) & (\num{0.023}) & (\num{0.052})\\
 
 [p-value] & {}[\num{0.931}] & {}[\num{0.913}] & {}[\num{0.160}]\\
\\
N Obs & {4,058,758} & {2,552,777} & {4,058,758}\\
    \hline 
    \hline \\[-1.8ex] 
    \end{tabular} 
    \begin{tablenotes}
    \item \footnotesize{Notes: The table presents the estimation of $\beta$ from equation \ref{res:main_results} for the following outcomes: indicator for women, indicator for Black/Native and age at admission. I multiply the coefficient by minus one, therefore we can see the coefficients as a decrease of 100\% of the weighted hiring measure. The recession between 2014-2016 reduced the weighted hiring measure by 30\%. Standard errors are clusteres at both the major and state levels (\emph{two-way clustering}). All regressions include fixed effects for program, major-admission time, and fall semester.}
    \end{tablenotes}
    \end{threeparttable}
    \end{adjustbox}
    \end{table} 
\begin{table}[!ht] \centering 
    \begin{adjustbox}{max width = \textwidth, width = \textwidth, center}
    \begin{threeparttable}
    \caption{Robustness Unemployment Rate}\label{res:robustness_unemp}
    \begin{tabular}{lcccccc} 
    \\[-1.8ex]\hline 
    \hline \\[-1.8ex]
    \multicolumn{1}{r}{\emph{Outcome:}} & \multicolumn{6}{c}{\emph{On-time graduation}}\\
    \addlinespace
    & (1) & (2) & (3) & (4) & (5) & (6) \\[.8ex] \hline \\[-.8ex]
    \addlinespace
    Unemployment & \num{0.024} &  &  &  &  & \\
 & (\num{0.164}) &  &  &  &  & \\
 & {}[\num{0.886}] &  &  &  &  & \\
 \\
Unemployment x Public &  & \num{-0.470} & \num{-0.479} & \num{-0.483} & \num{-0.382} & \num{-0.213}\\
 &  & (\num{0.301}) & (\num{0.268}) & (\num{0.285}) & (\num{0.356}) & (\num{0.359})\\
 &  & {}[\num{0.131}] & {}[\num{0.086}] & {}[\num{0.102}] & {}[\num{0.294}] & {}[\num{0.559}]\\
 \\
Unemployment x Private &  & \num{0.141} & \num{0.341} & \num{0.396} & \num{-0.083} & \num{0.033}\\
 &  & (\num{0.239}) & (\num{0.213}) & (\num{0.240}) & (\num{0.297}) & (\num{0.295})\\
 &  & {}[\num{0.560}] & {}[\num{0.122}] & {}[\num{0.112}] & {}[\num{0.783}] & {}[\num{0.911}]\\ 
 \\
    \midrule
    N Obs & \num{4666216} & \num{4666216} & \num{4669658} & \num{4669658} & \num{4669658} & \num{4666216}\\
    \\
    p-value ($\beta_{\text{public}}=\beta_{\text{private}}$) & - & \{0.148\} & \{0.031\} & \{0.041\} & \{0.527\} & \{0.619\} \\ 
    \\
    Major-State FE &$\checkmark$ & $\checkmark$ &- & - & - & - \\
    Program FE & -&-& $\checkmark$& $\checkmark$& $\checkmark$ & $\checkmark$\\
    Demographics & -& - & $\checkmark$& $\checkmark$&$\checkmark$ & $\checkmark$\\
    \addlinespace
    \multirow{2}{*}{Time Trend} & \multirow{2}{*}{Quadratic} & \multirow{2}{*}{Quadratic} & \multirow{2}{*}{Quadratic} & \multirow{2}{*}{Quadratic} & Admission & Major-Admission \\
    &&&&&Time FE & Time FE\\
    \hline 
    \hline \\[-1.8ex] 
    \end{tabular} 
    \begin{tablenotes}
    \item \footnotesize{Notes: The table presents the estimation of $\beta$ from equation \ref{res:main_results}. The coefficients can be interpreted as the effects in on-time graduation rate caused by an increase of 100 percentage points in the unemployment rate. The first column presents the overall results. In columns 2-6, I interact all variables with an indicator of whether the students belong to a private or public institution. Columns 2-6 differ in the set of control variables included in each specification. Standard errors are clustered at the state level. The p-value of the test whether the effect for public and private institutions are the same is provided for each specification. FE stands for fixed effects.}
    \end{tablenotes}
    \end{threeparttable}
    \end{adjustbox}
\end{table}

\end{document}